\documentclass[prd,preprint,floatfix,eqsecnum,nofootinbib,12pt]{revtex4}
\usepackage{hyperref}

\usepackage{amssymb,amsmath,amsthm,graphicx,ulem,slashed}
\pdfoutput=1
\usepackage{graphicx,subfigure}

\usepackage{ulem}
\usepackage{epsfig}
\usepackage{amsfonts}
\usepackage[usenames]{color}
\usepackage[letterpaper,left=2cm,right=2cm,top=2.32cm,bottom=2.32cm]{geometry}
\usepackage[T1]{fontenc}
\usepackage{verbatim}

\def\be{\begin{equation}}
\def\ee{\end{equation}}

\newcommand {\cN}{{\cal N}}

\def\b{\beta}

\def\e{\epsilon}

\def\g{\gamma}

\def\s{\sigma}

\def\ri{{\rm i}}

\newcommand{\pa}{\partial}                

\newcommand{\bea}{\begin{eqnarray}}
\newcommand{\eea}{\end{eqnarray}}
\newcommand{\non}{\nonumber}
\newcommand{\ba}{\begin{array}}
\newcommand{\ea}{\end{array}}

\def\double #1{#1{\hbox{\kern-2pt $#1$}}}

\newcommand{\de}{{\nabla}}

\newcommand{\bsubeq}{\begin{subequations}}
\newcommand{\esubeq}{\end{subequations}}

\usepackage{cleveref}

\begin{document}

\title{{\LARGE {Spinors in Supersymmetric dS/CFT} }}

\author{Thomas Hertog$^{a}$}
\email{thomas.hertog@kuleuven.be}
\author{Gabriele Tartaglino-Mazzucchelli$^{a,b,c}$}
\email{g.tartaglino-mazzucchelli@uq.edu.au}
\author{Gerben Venken$^{a}$}
\email{gerben.venken@kuleuven.be}
\affiliation{$~^{a}$Institute for Theoretical Physics, KU Leuven, Celestijnenlaan 200D, 3001 Leuven, Belgium
%}
%\affiliation{
\\
$~^{b}$Albert Einstein Center for Fundamental Physics,
Institute for Theoretical Physics,
University of Bern, Sidlerstrasse 5, CH-3012 Bern, Switzerland
\\
$~^{c}$School of Mathematics and Physics, 
University of Queensland St Lucia, Brisbane, 
Queensland 4072, Australia
}

\begin{abstract}

We study fermionic bulk fields in the dS/CFT dualities relating ${\cal N}=2$ supersymmetric Euclidean vector models with reversed spin-statistics in three dimensions to supersymmetric Vasiliev theories in four-dimensional de Sitter space. These dualities specify the Hartle - Hawking wave function in terms of the partition function of deformations of the vector models. We evaluate this wave function in homogeneous minisuperspace models consisting of supersymmetry-breaking combinations of a half-integer spin field with either a scalar, a pseudoscalar or a metric squashing. The wave function appears to be well-behaved and globally peaked at or near the supersymmetric de Sitter vacuum, with a low amplitude for large deformations. Its behavior in the semiclassical limit qualitatively agrees with earlier bulk computations both for massless and massive fermionic fields. 

\end{abstract}

\maketitle

\tableofcontents

\newpage

%%%%%%%%%%%%%%%%
\section{Introduction}
%%%%%%%%%%%%%%%%

Gauge-gravity duality with de Sitter (dS) boundary conditions \cite{Hull:1998vg,Balasubramanian2001,Strominger:2001pn} has proved to be a fruitful route to elucidate the status of de Sitter space in string theory and to put cosmology on firm theoretical ground. In its most ambitious and fundamental form, dS/CFT conjectures that the partition functions of certain deformations of three dimensional Euclidean CFTs yield a precise formulation of the Hartle-Hawking wave function of the universe \cite{Hartle:1983ai}. Schematically and in the large three-volume regime the proposed dual form of the wave function reads 
\be
\Psi_{HH} [h_{ij}, A_s]= Z_{QFT}[\tilde h_{ij}, J_s] \exp(\ri S_{st}[h_{ij}, A_s]/\hbar)   \ .
\label{dSCFT}
\ee
Here $A_s$ stands for matter configurations of spin $s$ and $h_{ij}$ is the three-geometry of the spacelike surface $\Sigma$ on which $\Psi$ is evaluated. In this paper we take the latter to be topologically a three-sphere. The sources $(\tilde h_{ij}, J_s)$ in \eqref{dSCFT} are conformally related to the arguments $(h_{ij}, A_s)$ of the wave 
function, and $S_{st}$ are the usual surface terms.

It is a central open question what class of deformed CFTs in \eqref{dSCFT} specifies a well-defined, normalizable wave function. Euclidean AdS/CFT provides a starting point to study this since its generalization to complex relevant deformations of CFTs implies a semiclassical realisation of dS/CFT \cite{Maldacena:2002vr,Harlow2011,Maldacena2011,Hertog2011} that is  possibly exact in Vasiliev gravity in dS \cite{Anninos:2011ui}. It has been suggested indeed that Euclidean AdS and Lorentzian dS, and their duals, can be viewed as two real domains of a single complexified theory \cite{Maldacena:2002vr,Hull:1998vg,Dijkgraaf:2016lym,Skenderis:2007sm,Bergshoeff:2007cg,Hartle2012b}. 

An interesting point in this respect is that in dS/CFT the Euclidean duals are never Wick rotated to the Lorentzian. It is therefore misguided to criticize dS/CFT on the grounds that the duals are not unitary\footnote{One may argue that reflection positivity is the relevant notion for Euclidean theories and that the CFTs in dS/CFT are not reflection positive. However, reflection positivity is similarly not an interesting property for a Euclidean theory by itself. A reflection positive theory is usually only reflection positive along a single preferred direction. Reflection positivity is therefore relevant when one intends to Wick rotate this direction into a time direction, since this guarantees that the resulting Lorentzian theory will be unitary. But we do not Wick rotate in dS/CFT, and time can be viewed as emerging holographically in the bulk. Therefore, there is no natural boundary direction along which one should impose reflection positivity.}. This is an important conceptual difference with AdS/CFT. Of course, if the wave function is well-behaved then it predicts unitary time evolution in the bulk at the level of quantum field theory in each of the asymptotically classical spacetime backgrounds it describes. It is clearly important to better understand what this entails when it comes to the dual.

The case of higher-spin (HS) gravity provides an interesting toy model to explore these questions, since the duals are vector models for which the partition functions can be evaluated explicitly for a range 
of deformations \cite{Anninos:2012ft,Anninos:2013rza,Bobev:2016sap,Conti:2017pqc,Hawking:2017wrd}. The Vasiliev HS theory has massive scalars and an infinite tower of massless gauge fields of increasing spin \cite{Vasiliev:1990en}. The duals have conserved currents for the same symmetries \cite{Klebanov:2002ja,Giombi:2016ejx}. Deforming the boundary theory action with a conserved current $J_s$ corresponds to turning on the spin-$s$ field $A_s$. Calculations of the partition function with homogeneous scalar and spin-2 
deformations in the $Sp(N)$ vector model, dual to the minimal Vasiliev theory in dS \cite{Anninos:2011ui}, have provided some evidence that dS/CFT yields a well-defined wave function and in particular one which is better behaved than the usual semiclassical Hartle-Hawking wave function in Einstein gravity.

In recent work \cite{Hertog:2017ymy} we put forward a supersymmetric generalization of these HS dualities in dS.\footnote{It is often argued that unbroken supersymmetry and dS space do not go together (see e.g. \cite{Pilch:1984aw,Lukierski:1984it,Witten:2001kn}) because in dS space there is no positive conserved quantity whereas supersymmetry would allow one to construct one. However supersymmetric HS gravity theories in de Sitter circumvent this problem since the Hermitian conjugate in the theories in \cite{Sezgin:2012ag} is an anti-involution \cite{Hertog:2017ymy}. In a similar spirit one may object that de Sitter space `has a temperature' and therefore cannot be supersymmetric. However, as Gibbons and Hawking \cite{Gibbons:1977mu} already pointed out, the temperature arises only from an observer's perspective. The wave function that dS/CFT computes is a function over global configurations and may itself be a SUSY invariant pure state. To obtain a physical description relevant to local observers, one should trace over the degrees of freedom outside their subjective horizon. This produces a SUSY breaking thermal density matrix. Essentially the same argument has been given in \cite{Anous:2014lia} in the context of superconformal field theory on a de Sitter background.} The bulk theories involved are the supersymmetric extensions of Vasiliev theory described in \cite{Sezgin:2012ag}. On the boundary side we constructed new ${\cal N}=2$ supersymmetric extensions of the three-dimensional $Sp(N)$ models. We then related these to the theories of Sezgin and Sundel, thereby establishing a supersymmetric gauge-gravity duality with de Sitter boundary conditions. We evaluated the partition function of these supersymmetric extensions of the free $Sp(N)$ model with homogeneous scalar, vector and spin-2 deformations that preserve supersymmetry. The duality \eqref{dSCFT} conjectures that these partition functions specify the Hartle-Hawking wave function in a supersymmetric minisuperspace consisting of anisotropic deformations of de Sitter space with scalar and vector matter. We found the wave function is globally peaked at the undeformed de Sitter space with a low amplitude for strong deformations, indicating that supersymmetric de Sitter space in higher-spin gravity is stable and has no ghosts\footnote{As an aside we note that supermultiplets with flipped spin-statistics have also appeared in a different context in CFTs in \cite{Buican:2017rya,Buican:2019huq} where non-unitary 4D theories are related to unitary 2D theories. This gives an example where an apparently non-unitary theory encapsulates the data of a unitary theory. The non-unitary theory carries in fact a hidden notion of unitarity, and this with precisely the same ingredients as the `non-unitarity' theories in dS/CFT, suggesting it is reasonable indeed to expect that bulk unitarity could arise holographically in dS/CFT.}. 

In this paper we initiate the study of fermions in dS/CFT. In particular we compute the contribution from bulk fermions to the Hartle-Hawking ground state wave function $\Psi_{HH}$ in these supersymmetric HS theories. The properties of superpartner fermions are especially interesting since in higher spin theory, fermionic fields can essentially only arise as superpartners to bosonic fields.  We find that, within the analysis performed so far, the existence of a stable supersymmetric de Sitter vacuum - our fundamental conclusion in \cite{Hertog:2017ymy} - remains unchanged with the inclusion of these fermionic fields. As before we perform all calculations in the conjectured dual and use the duality \eqref{dSCFT} to obtain the Hartle-Hawking wave function. Purely computationally, the dS/CFT dictionary \eqref{dSCFT} applies directly to fermionic fields. However, the interpretation of $\Psi_{HH}$ in the presence of fermionic fields requires a certain care as detailed long ago in \cite{DEath:1986lxx,DEath:1996ejf}. 

The paper is organised as follows. In \cref{secwave functionfermion} we begin with a general discussion of fermionic bulk fields in the wave function of the universe and in the context of dS/CFT. This discussion is independent of the specifics of our model and the presence of supersymmetry. In \cref{sec:deform}, we summarise the model of supersymmetric higher spin dS/CFT we constructed in \cite{Hertog:2017ymy}, focussing especially on how the fermionic bulk fields enter. In \cref{Spin1/2}, we deform the boundary theory to turn on a spin-$1/2$ fermion in the bulk, while keeping all other bulk fields turned off. We study how the bulk wave function and observables respond to this fermionic field. In \cref{secinterplay}, we repeat this process, now simultaneously turning on both a fermionic and a bosonic bulk field. We consider scalar, pseudoscalar and metric bulk fields. This allows us to study the interplay between fermionic and bosonic bulk fields. We see no sign of instability of the bulk vacuum coming from this interplay. In \cref{sec:discussion}, we first discuss the behaviour of $\Psi_{HH}$ in the presence of bulk fermions in our model and we then conclude with a more general discussion of dS/CFT in string theory.

%%%%%%%%%%%%%%%%
\section{Bulk fermions and the wave function}
\label{secwave functionfermion}
%%%%%%%%%%%%%%%%

As we have described in \cite{Hertog:2017ymy}, 
half-integer spin CFT sources  enter into the Hartle-Hawking wave function and the duality \cref{dSCFT} in a way that is formally completely analogous to the bulk bosons. However, on a conceptual level, the interpretation of these fields in the wave function deserves some extra attention. Here we discuss in general terms how spinorial bulk de Sitter fields enter into the dS/CFT correspondence, not limiting ourselves to our particular model.

In the action of a QFT, fermionic fields are often described using Grassmann variables. 
But, it is important to remember that, just as quantised bosons are not simply commuting variables, quantised fermions are not Grassmann variables. The Grassmann variables simply provide a convenient way to describe the fermions pre-quantisation. Let us briefly review the canonical quantisation of fermions. Our treatment here will be rather brief and qualitative. A more detailed discussion of the quantization of a free fermion in de Sitter space and the properties of the $\breve{\psi}^{A'}$ conjugate spinors can be found in \cref{freespinorapp}. To keep our presentation concise, we will sometimes suppress constant factors and indices when not relevant. A more complete discussion can be found in \cite{DEath:1986lxx,DEath:1996ejf}. Consider the bulk spinor $\psi(x)$ in four-dimensional de Sitter space at some time-slice. The spatial geometry on this time-slice is an $S^3$ and the spinors can be expanded in spherical harmonic (commuting) spinors.\footnote{On a deformed geometry the same line of reasoning goes through, with the harmonics of the deformed geometry.} The three-sphere harmonics are $\rho^{n p}_A$, $\sigma^{n p}_{A'}$ with positive frequencies and their conjugates $\overline{\rho}^{n p}_{A'}$, $\overline{\sigma}^{n p}_A$ with negative frequencies. Here, $A$ is the bulk spinor index, $n=0,1,2,..$ tracks the eigenvalue under the harmonic equation and $p=1,2,..,(n+1)(n+2)$ tracks the degeneracy for given $n$. Ignoring  the degeneracy (which will be broken anyway when we deform the geometry), one can expand as
\bsubeq
\begin{align}
\psi_A &\sim \sum_n \big( s_n \rho^n_A + \breve{t}_n \overline{\sigma}^n_A \big) \,, \\
\breve{\psi}_{A'} &\sim \sum_n \big( \breve{s}_n \overline{\rho}^n_{A'} + t_n \sigma^n_{A'} \big) \,.
\end{align}
\esubeq
If one is describing a classical and commuting spinor function; $s$, $t$ and their breves should be constant coefficients.\footnote{If one wants to describe how $\psi$ evolves in time rather than restricting to one time-slice, these coefficients would become time-dependent.} However, we wish to quantise $\psi$ as a fermion. Then as usual, we want these coefficients to essentially play the role of creation and annihilation operators. This means that we should impose the anticommutation relations
\bsubeq
\begin{align}
\label{quantization1}
\{ s_n, t_m \} &= \{ s_n, \breve{t}_m \}= \{ \breve{s}_n, t_m \} = \{ \breve{s}_n, \breve{t}_m \}= 0 \,, \\
\{ s_n, \breve{s}_m \} &= \{ t_n, \breve{t}_m \} = \delta_{n m} \label{anncreat} \,.
\end{align}
\esubeq

As pointed out in \cite{Atiyah:1975jf,DEath:1986lxx,DEath:1996ejf}, one should take the wave function on the future boundary surface to depend only on the positive frequency modes. This is convenient as the fermionic dependence of the wave function is then only a dependence on anticommuting Grassmann variables $s$ and $t$. If the negative frequency modes had entered as well, we would have got variables with nontrivial anticommutation relations.

It is then clear how, for example, we should interpret the $J_{1/2}$ source of the boundary CFT related to a spin-$1/2$ bulk field, which enters into the wave function as Grassmann valued spinors, rather than fully quantised fermionic fields. In the boundary CFT, we can turn on a separate source term
\begin{equation}
J_{1/2 \,, n} \mathcal{O}_{1/2}
\end{equation}
for each three-sphere harmonic. In our specific model, $J_{1/2}$ is related to a background gaugino field $\lambda$ in the CFT and $\mathcal{O}_{1/2}$ to a combination of dynamical CFT scalar and spinor, $\overline{\chi} \varphi$ and we will adopt this notation from here on for future convenience.

We then have a separate source term
\begin{equation}
 \lambda_n (x) \overline{\chi} \varphi + \chi \breve{\lambda}_n (x) \overline{\varphi}
\end{equation}
for each three-sphere harmonic. We associate $\lambda_n$ in the boundary theory with $s_n \rho^n_A$ (no Einstein summation) in the bulk wave function and we associate $\tilde{\lambda}_n$ with $t_n \sigma^n_{A'}$.\footnote{Or the association is the other way, that is arbitrary.}

The purpose of the CFT in dS/CFT is to compute the bulk wave function $\Psi$. So, our sources are just Grassmann variables and from the path integral CFT perspective, we need not concern ourselves with the nontrivial commutation relations \ref{anncreat}. Of course, once we have our bulk wave function, to be able to interpret it and study its properties, we need a notion of conjugate wave function and a definition of inner product between two wave functions with fermionic fields. Here, $\tilde{s}$, $\tilde{t}$ and the nontrivial commutation relations \ref{anncreat} will come into play. The correct description of a second quantised fermionic field in a wave function formalism  has long been known. The classic references are \cite{Berezin:1966nc,Faddeev:1980be} and a discussion in the context of quantum cosmology can be found in \cite{DEath:1986lxx,DEath:1996ejf}. The point is that the inner product should be defined such that wave functions with different fermion states occupied should be orthonormal. This is what one naturally expects and how things also work in the, for QFT more conventional, description in terms of state vectors in a Hilbert space. One can then expand the wave function in terms of the Grassmann variables, where due to their Grassmann nature, each $\lambda_n$, $\tilde{\lambda}_n$ can appear only twice at most in one term, but one can of course have many different $\lambda_n$ with different $n$ in one term\footnote{The wave function could in principle also include terms for instance of the type $(\lambda \lambda)$. The dynamics of our specific theory will ensure we only have $(\tilde{\lambda}\lambda)$ terms appearing. The physical picture here is that $\tilde{\lambda}$ is associated to particle creation and $\lambda$ to antiparticle creation. If the dynamics of the theory produces particles and antiparticles in pairs, one then expects only terms of the form $(\tilde{\lambda}\lambda)$.}.
 \begin{equation}
\label{definitionfermionwave function}
\Psi = \Psi_{\text{Bosonic}} +  \Psi_0 (\tilde{\lambda}_0 \lambda_0) + \frac{1}{2} \Psi_{00} (\tilde{\lambda}_0 \lambda_0)^2 + \Psi_1 (\tilde{\lambda}_1 \lambda_1) + \Psi_{01} (\tilde{\lambda}_1 \lambda_1)\ (\tilde{\lambda}_0 \lambda_0 ) + \frac{1}{2} \Psi_{001} (\tilde{\lambda}_1 \lambda_1)\ (\tilde{\lambda}_0 \lambda_0 )^2 + ..
\end{equation}
The $\Psi_{i_1 \cdots i_k}$ only depend on the bosonic fields and the inner product between two bulk wave functions is given by
\begin{equation}
(\Psi ,\Phi ) = (\Psi_{\text{Bosonic}} ,\Phi_{\text{Bosonic}} )_B + (\Psi_0 ,\Phi_0 )_B + (\Psi_{00} ,\Phi_{00} )_B + (\Psi_1 ,\Phi_1 )_B + (\Psi_{01} ,\Phi_{01} )_B + (\Psi_{001} ,\Phi_{001} )_B + .. \, \,,
\label{innerproduct}
\end{equation}
where $(\, , \, )_B$ refers to the inner product over the bosonic bulk fields.
For simplicity, we will only study the constant spinor $\lambda_0$ harmonic in this paper and will refer to it simply as $\lambda$ in the future, but all other harmonics enter in an analogous way.

An interesting quantity to compute is the number operator for the fermion zero mode, $N_F$. When this operator acts on a state, it gives the number of $\tilde{\lambda} \lambda$ in that state.  So, it acts as
\begin{equation}
N_F \Psi = N_F (\Psi_B + \Psi_0 + \Psi_{00}) = 0 \Psi_B + 1 \Psi_0 + 2 \Psi_{00}
\end{equation}
on a state $\Psi$.
Its expectation value is given by
\begin{equation}
< N_F > = \frac{(\Psi, N_F \Psi)}{(\Psi, \Psi)} =\frac{(\Psi_0 , \Psi_0 )_B + 2 (\Psi_{00}, \Psi_{00})_B}{(\Psi_{\text{bosonic}}, \Psi_{\text{bosonic}})_B + (\Psi_0 , \Psi_0 )_B + (\Psi_{00}, \Psi_{00})_B} \,.
\end{equation}
This takes a value between zero and two. The denominator in this expression serves to normalize the state. When its value is low, empty de Sitter space is preferred, for a high value, a state with bulk fermions present is preferred. For a massless free fermion in de Sitter space in the Hartle-Hawking state, a bulk computation by \cite{DEath:1986lxx,DEath:1996ejf} has shown that $<N_F>=0$.
In principle, this is a perfectly valid observable to consider. 
But, when we consider a minisuperspace that involves both bosonic and fermionic deformations, we run into a practical issue. It is in general unclear what the precise definition of the bosonic inner product $(\ ,\ )_B$ should be, preventing us from actually evaluating $< N_F >$. To deal with this, we can instead compute the local value of $< N_F >$ at a given boundary value $B$ of the bosonic bulk fields
\begin{equation}
< N_F >[B] = \frac{\Psi^*_0[B] \Psi_0[B] + 2 \Psi^*_{00}[B] \Psi_{00}[B]}{\Psi^*_{\text{bosonic}}[B] \Psi_{\text{bosonic}}[B] + \Psi^*_0[B] \Psi_0[B] + \Psi^*_{00}[B] \Psi_{00}[B]} \,,
\label{NFB}
\end{equation}
where for the wave functions that depend only on the bosonic fields, the conjugate wave function is simply the complex conjugate.
Note that we have done something unusual in the denominator of this expression. The denominator serves to normalise our expression and normally we should normalise against the full value of the wave function $(\Psi, \Psi)$. However, here we are normalising against the value of the wave function at the given bosonic deformation $B$. This means that we effectively consider the wavefunction of a state where we have already fixed bosonic fields to a certain value.
This ensures we have values between zero and two, which we will see leads to a clearer and more interesting analysis. We should keep in mind that we are no longer truly computing something which when we integrate it over $B$ with the correct measure will give us $< N_F >$ for the holographic no-boundary state. Instead, we are saying: suppose we have already imposed that the bosonic bulk fields take value $B$ on the future boundary. After making this imposition on our state, how do the fermionic spinor bulk fields respond to the bosonic bulk fields as an imposed background? Because of this, we need to normalise the wave function at given $B$. As we will see, this will lead to an interesting analysis in our model.

Another interesting observable to consider is $<\tilde{\lambda} \lambda>$. Clearly, its value is given by
\begin{equation}
<\tilde{\lambda} \lambda> = (\Psi_{\text{bosonic}}, \Psi_0)_B +  (\Psi_0, \Psi_{\text{bosonic}})_B + (\Psi_0, \Psi_{00})_B + (\Psi_{00},\Psi_0)_B \,.
\end{equation}
Again, it would be problematic to evaluate $(~,~)_B$ in a minisuperspace where we also consider the bosonic bulk fields. Instead, we can again consider it at a given value of the bosonic bulk fields $B$,
\begin{equation}
\label{spinorvevdef}
<\tilde{\lambda} \lambda>[B] = \Psi^*_{\text{bosonic}}[B] \Psi_0[B] +  \Psi^*_0[B] \Psi_{\text{bosonic}}[B] + \Psi^*_0[B] \Psi_{00}[B] + \Psi^*_{00}[B] \Psi_0[B] \,.
\end{equation}
Note that in this case we are not imposing normalisation against the wave function at $B$. This is because we want to keep the expression in a form where we could in principle integrate over $B$ with the correct measure to obtain $<\tilde{\lambda} \lambda>$ for the holographic no-boundary state. 
We will see that in our model, despite not knowing the exact measure to use, we will still be able to draw some interesting conclusions for $<\tilde{\lambda} \lambda>$ of the no-boundary state from the results we obtain. As the wavefunction is not normalized, we can interpret how $<\tilde{\lambda} \lambda>$ changes with $B$, but the actual numerical value of  $<\tilde{\lambda} \lambda>$ cannot be trusted as there might be a rescaling by a constant factor. In particular, wavefunctions computed at different $N$ in the CFT might have different normalization rescalings, resulting in different rescalings of  $<\tilde{\lambda} \lambda>$.

One can construct other observables related to fermionic bulk fields analogously to the ones we have discussed. It is clear from the form of \cref{definitionfermionwave function} that e.g. $<\lambda>=0$.

%%%%%%%%%%%%%%%%
\section{Supersymmetric Vector Models and Duality}
\label{sec:deform}
%%%%%%%%%%%%%%%%

First, let us review the work done in \cite{Hertog:2017ymy}, focussing especially on the specific results we need here.
The simplest bulk de Sitter higher spin theory consists of a conformally massive scalar of mass $m^2 = 2 / l^2$ and an infinite tower of massless fields of every even spin starting at two. The CFT dual to the simplest de Sitter higher spin theory is the free $Sp(N)$ vector model, consisting of anticommuting scalars \cite{Anninos:2011ui}. In \cite{Hertog:2017ymy}, an $\mathcal{N}=2$ supersymmetric extension of the $Sp(N)$ model, which we called the $\mathcal{N}=2$ $U(-N)$ model,  was constructed and argued to be dual to the $\mathcal{N}=2$ supersymmetric de Sitter higher spin theory constructed in \cite{Sezgin:2012ag}. The $\mathcal{N}=2$ supersymmetry extends the spectrum of the bulk higher spin theory such that there are no longer just even spins present, instead fields of every possible spin arise. That is, there are a scalar and pseudoscalar, conformally massive $m^2 = 2 / l^2$, followed by massless fields of every spin $s= 1/2, 1, 3/2, 2, 5/2,.., \infty$. This entire tower belong to a representation of the $ho(1,1|4,1)$ higher spin symmetry. Restricted to the  sector of spin up to and including two, which we will be especially interested in, this yields an $OSp(2|2,2)$ supersymmetry subgroup with eight real supercharges. The exact full bulk spectrum is in one-to-one correspondence with boundary single trace conserved supercurrents and can in this way be elegantly captured in the language of $3D$ $\mathcal{N}=2$ superfields as described in \cite{Hertog:2017ymy}.

The action of the undeformed supersymmetric $U(-N)$ model on a round $S^3$ boundary sphere of radius 
$l$ is
\begin{align} 
    S_0 = \int d^3x \sqrt{h} \left[ \partial_\mu \tilde{\varphi}_i \partial^\mu \varphi^i 
    + \frac{3}{4 l^2} \tilde{\varphi}_i \varphi^i 
    +\ri \tilde{\chi}_i \slashed{\nabla} \chi^i 
    - \tilde{G}_i G^i\right]\text{.}
\label{chiralS3}
\end{align}

Here, all fields have had their spin-statistics flipped. That is, $\varphi^i$ and $G^i$ are anticommuting scalars, $\chi^i$ is a commuting spinor, analogously for the ``tilde'' fields and $i=1,\dots,N$. We follow the same notation as \cite{Hertog:2017ymy}, convenient for Euclidean $3D$ $\mathcal{N}=2$ supersymmetric theories. Importantly, this means that the ``tilde'' fields are the ``would-be'' conjugate fields. These are fields that are allowed to be taken independent from the fields without tilde while still preserving supersymmetry. However, were we working in Lorentzian signature instead, they would be required to be hermitian conjugates to the fields without tilde. For the dynamical fields appearing in \cref{chiralS3}, we will indeed take the fields with and without tilde to be independent. On the background field 
which we will introduce in the following and that couple to the higher-spin currents constructed out of the dynamical
fields in \eqref{chiralS3}, 
on the other hand, we will impose additional reality conditions stemming from the reality conditions of their bulk dual fields.
We will comment more on this point at the end of the section.

The action \eqref{chiralS3} is invariant under the supersymmetry transformations of the matter chiral multiplet
\bsubeq\bea
 \delta \varphi^i &=& \sqrt{2} \e \chi^i
 ~, 
 \\
\delta \chi^i &=& \sqrt{2} \e G^i 
- \sqrt{2} \ri \gamma^\mu \tilde{\e} \partial_\mu \varphi^i 
-\frac{1}{\sqrt{2} l}\tilde{\e} \varphi^i
~, 
\\  
\delta G^i &=& - \sqrt{2} \ri \tilde{\e} \gamma^\mu \nabla_\mu \chi^i
~,
\eea\esubeq
and similarly for the ``tilde'' fields,
where  ($\e,\,\tilde{\e})$ and $\de_\mu$ are respectively Killing spinors and covariant derivatives on the three-sphere
$S^3$, see e.g. \cite{Willett:2016adv} for more detail.\footnote{As in \cite{Hertog:2017ymy} we
use the notation and conventions of \cite{Willett:2016adv} up to name redefinition of the fields of the matter sector
and of the Killing spinors.
Note also that, compared to the result of \cite{Willett:2016adv}, there are some different signs due to the 
reversed statistic of the fields in the chiral multiplet of the $U(-N)$ model.}

In \cite{Hertog:2017ymy} we considered a class of scalar deformations of the previous action that arise from the 
supersymmetric coupling of the $U(-N)$ model to source spin-zero currents.
In practice, these couplings, that amount to particular mass deformations once one restricts to constant deformations,
arise as a subsector of the $U(-N)$ model coupled to a background Abelian supersymmetric vector multiplet.
The main scope of this paper is to analyse deformations induced by fermionic sources that typically manifestly break supersymmetry and the isometries of $S^3$. 
The simplest of these deformations
arise from the coupling of the spin-1/2 current of the supersymmetric $U(-N)$ model  to a background gaugino. Let us show explicitly how this term arises.

First of all, we remind the reader that an Abelian vector multiplet on the sphere is described by the following fields: 
the gauge connection $A_\mu$ with field strength $F_{\mu\nu}=\pa_{[\mu}A_{\nu]}$; two gaugini fermionic fields
$\lambda$ and $\tilde{\lambda}$; a scalar and pseudoscalar fields $\varsigma$ and $D$.
Their supersymmetry transformations on the sphere are given by 
\bsubeq\bea
\label{susyvar}
\delta A_\mu &=& -\ri (\e \gamma_\mu \tilde{\lambda} + \tilde{\e} \gamma_\mu \lambda )
~,
\\
\delta \varsigma &=& - \e \tilde{\lambda} + \tilde{\e} \lambda 
~,
\\
\delta \lambda 
&=& 
\Big(\ri  D 
- \frac{\ri}{2} \epsilon^{\mu \nu \rho} \gamma_\rho  F_{\mu \nu} 
- \ri \gamma^\mu  \de_\mu \varsigma  
-\frac{1}{l}\varsigma \Big) \e
~,
\\
\delta \tilde{\lambda} &=&
 \Big(- \ri D
 - \frac{\ri}{2} \epsilon^{\mu \nu\rho} \gamma_\rho F_{\mu \nu} 
 + \ri \gamma^\mu \de_\mu \varsigma 
  + \frac{1}{l} \varsigma \Big) \tilde{\e} 
 ~,
 \\
\delta D 
&=& 
 \e \gamma^\mu \de_\mu \tilde{\lambda} 
 - \tilde{\e} \gamma^\mu \de_\mu \lambda 
 + \frac{1}{3} ( \nabla_\mu \e \gamma^\mu \tilde{\lambda} - \nabla_\mu \tilde{\e} \gamma^\mu \lambda)
 ~.
 \eea
\esubeq
Before continuing it is important to underline that these fields have the standard spin-statistics.

The coupling of the $U(-N)$ model
to a background gauge supermultiplet amounts to adding to the action \eqref{chiralS3} the following 
interaction terms
\bea
S_{gauge-int} &=&
\int d^3x \sqrt{h} \Big{[}
 \tilde{\varphi}_i ( \varsigma^2 + D) \varphi^i  
- \ri \varsigma \tilde{\chi}_i\chi^i 
 + \sqrt{2} \ri (\tilde{\varphi}_i \lambda \chi^i - \tilde{\chi}_i \tilde{\lambda} \varphi^i )
 \Big{]}
 ~.
 \label{chiral-S3-gauge}
\eea
The action $S_0+S_{gauge-int}$ is invariant under supersymmetry transformations provided that the 
transformation for the matter multiplet is modified to
\bsubeq
\bea
\delta \varphi^i &=& \sqrt{2} \e \chi^i
~,\\
\delta \chi^i &=& 
\sqrt{2} \e G^i 
- \sqrt{2} \ri \gamma^\mu \tilde{\e} D_\mu \varphi^i
+ \sqrt{2} \ri \varsigma \tilde{\e}   \varphi^i
-\frac{1}{\sqrt{2} l}\tilde{\e} \varphi^i
~, 
\\
\delta G^i &=&
 - \sqrt{2} \ri \tilde{\e} \gamma^\mu D_\mu \chi^i 
 -\sqrt{2} \ri \varsigma \tilde{\e} \chi^i 
 + 2 \ri  \tilde{\e} \tilde{\lambda} \varphi^i
 ~,
 \eea
 \esubeq
where the $D_\mu$ derivatives are gauge covariant.

In \cite{Hertog:2017ymy} we noticed that an interesting deformation that amount to giving 
a supersymmetric mass to the chiral multiplets is based on choosing
a constant BPS configuration for the scalar fields $\varsigma$ and $D$ in the vector multiplet
where $D=-\ri \varsigma/l=const$. 
Substituting this condition in \eqref{chiral-S3-gauge} yields the following mass deformation,
\begin{equation}
\label{imaginarymassdeform}
    \mathcal{L}_{mass} = \left(\varsigma^2 - \ri \frac{\varsigma}{l} \right) \tilde{\varphi}_i \varphi^i -\ri \varsigma \tilde{\chi}_i \chi^i
    \text{ ,}
\end{equation}
where $l$ is the radius of the sphere. Further, we also studied a BPS combination of a spin-1 and spin-2 deformation in our earlier work. Here, results from supersymmetric localisation allowed us to compute the partition function exactly. Since we constructed a supersymmetric dS/CFT correspondence in our earlier work, it seemed natural to study BPS configurations there; both to show the simplifying power of supersymmetry to our wave function computations and because we believed that if supersymmetry would introduce an instability in our theory, supersymmetric deformations would be the most probable culprit. We found in \cite{Hertog:2017ymy} that our vacuum is stable under BPS deformations.

If we aim at turning on a spin-1/2 source on the boundary we should turn on the gauginos in the last two terms of
\eqref{chiral-S3-gauge}. This will be the main class of deformation we will study in this paper. Note that such a deformation will generically break all supersymmetries. We will compute the partition function and interpret it as the Hartle-Hawking wave function both for the spin-1/2 source by itself and for the combination of the spin-1/2 with a number of integer spin sources. Since the spin-1/2 source breaks supersymmetry, there is no longer a reason to consider the integer spin sources in would-be BPS configurations. It is also no longer possible to use results from localisation techniques 
to e.g. compute the spin-1 + spin-2 `BPS' deformation partition function exactly.

We want to make an important remark about what we (do not) mean by a source breaking supersymmetry. We mean that the CFT action with that source turned on is no longer supersymmetric. We do not mean that the bulk Hartle-Hawking state loses its supersymmetry. In fact, one should expect the full wave function, assuming one hypothetically fully computes it over all possible bulk field configuration, to be a supersymmetric ground state of the supersymmetric bulk theory. When one evaluates the inner product of the bulk wave function, one integrates over all bulk field configurations. From the CFT perspective, one then integrates over all configurations of the sources. These sources are described by background fields, which enter as supermultiplets. The integration over all source configurations then becomes a path integral over the `background fields', which have now become dynamical\footnote{In the sense that they enter into the path integral, there are no kinetic terms for these fields in the action.} by this procedure. The background fields for all spins enter in supermultiplets, which can be described in superspace 
using techniques of \cite{Kuzenko:2011xg,Kuzenko:2011rd} as has been done in \cite{Hertog:2017ymy}, making it manifest that supersymmetry is preserved. So, the full bulk wave function is manifestly supersymmetric. In practice, it is only possible to compute the wave function for a subset of all field configurations. The wave function evaluated over this subset might fail to be supersymmetric but this is merely an unfortunate side effect of restricting to the calculations one can in practice do. When one goes to the full configuration space, supersymmetry is always restored. This is entirely analogous to how a particular configuration that enters into the Hartle-Hawking wave function by itself does not preserve diffeomorphisms, but the wave function as a whole has a diffeomorphism-invariant inner product.

We still need to address the scaling of our sources with $N$. We can redefine our sources such that different factors of $N$ appear in front of them. This results in the corresponding bulk quantities having different scaling with $N$ in their $n$-point functions. When one discusses the bulk wave function for bosonic fields, this only results in a different overall factor of $N$ in the wave function and the issue need not be explicitly addressed. We shall see that with half-integer bulk fields, there is a qualitatively different behaviour for different values of $N$ and we need to make a physically sensible choice of scaling with $N$ for our sources. We demand that the bulk two-point functions produced by our rescaled sources have unit strength with regard to (do not scale with) $N$. In higher spin theories, this is achieved by scaling them such that there is a factor $1/\sqrt(N)$ in the boundary action in front of the source terms \cite{Sezgin:2002rt,Klebanov:2002ja,Leigh:2003gk,Sezgin:2003pt}. This is the scaling we will choose in the rest of the paper.

Loosely speaking, half-integer spin fields enter into higher spin theories `as superpartners' of the integer spin fields. While it is permissible to have more integer spin fields than half-integer spin fields, the reverse is not permitted
\cite{Sezgin:2012ag}. From the CFT perspective, this is intuitively clear by combinatorics. Conserved currents whose terms have either two boundary scalars or two boundary spinors yield integer spin fields in the bulk; whereas conserved currents whose terms have one boundary scalar and one boundary spinor produce half-integer spin bulk fields. Then, when there are as many boundary scalars as spinors, there are as many half-integer conserved currents as integer conserved currents, while in all other cases there are less half-integer conserved currents. So, our investigation here should be suggestive for spinors in generic, non necessarily supersymmetric, higher spin theories. If issues arise in our restricted case, one expects half-integer spin fields in de Sitter higher spin theories to be generically pathological.

In the following sections, we will consider turning on sources that can be written in new variables as
\begin{equation}
S = \int d^3 x \sqrt{h_{\text{def}}}
\left[ 
\begin{array}{c|c}
\tilde{\chi} &  \tilde{\varphi}
\end{array}
\right]
\left[
\begin{array}{c|c}
\ri \slashed{\nabla} + \frac{m}{\sqrt{N}} & - \sqrt{\frac{2}{N}}  \ri \tilde{\lambda} \\ \hline
\sqrt{\frac{2}{N}}  \ri \lambda &- \partial^2 + \frac{R}{8} + \frac{\sigma}{\sqrt{N}}
\end{array}
\right]
\left[
\begin{array}{c}
\chi \\ \hline
\varphi
\end{array}
\right]\text{,}
\end{equation}
where $m$ and $\sigma$ must be real, as discussed in \cite{Hertog:2017ymy}.
Here, $\sigma$ sources a scalar bulk field, $m$ a pseudoscalar bulk field, and we will also consider a squashing $h_{\text{def}}$ of the boundary metric.
Imposing that $m$ and $\sigma$ must be real imposes in terms of the variables used in \eqref{chiral-S3-gauge} that $D$ must be real and $\varsigma$ must be imaginary. This in turn imposes reality conditions on the other background fields through the supersymmetry variations \cref{susyvar}. In particular, one can see that $\tilde{\lambda}\lambda$ must be real.
When we compute $<\tilde{\lambda}\lambda>$ later, we shall see that it is indeed real, consistent with the reality conditions coming from supersymmetry. This is nontrivial as \cref{spinorvevdef} is not by definition real for an arbitrary theory.

%%%%%%%%%%%%%%%%
\section{Spin-1/2 deformation}
\label{Spin1/2}
%%%%%%%%%%%%%%%%

As discussed in \cref{sec:deform}, the spin-1/2 current deformation is given by adding
\begin{equation}
\sqrt{\frac{2}{N}}  \ri (\tilde{\varphi}_i \lambda \chi^i - \tilde{\chi}_i \tilde{\lambda} \varphi^i )
\end{equation}
to the Lagrangian. Constant spinors\footnote{Constant in the basis naturally constructed by recalling that $S^3$ is the $SU(2)$ group manifold, as e.g. in \cite{Willett:2016adv}.} exist on $S^3$ and in what follows, we will take $\lambda$ and $\tilde{\lambda}$ to be constant as discussed in \cref{secwave functionfermion}.

For the spin-1/2 deformation, the path-integral of the partition function is Gaussian, so an exact calculation of the partition function is in principle possible if we know the spectrum of the operators whose determinants we need. The action can be schematically written in the form
\begin{equation}
\label{gauginomatrixaction}
S = \int d^3 x \sqrt{h}
\left[ 
\begin{array}{c|c}
\tilde{\chi} &  \tilde{\varphi}
\end{array}
\right]
\left[
\begin{array}{c|c}
\ri \slashed{\nabla} & - \sqrt{\frac{2}{N}}  \ri \tilde{\lambda} \\ \hline
\sqrt{\frac{2}{N}}  \ri \lambda &- \partial^2 + 3/4
\end{array}
\right]
\left[
\begin{array}{c}
\chi \\ \hline
\varphi
\end{array}
\right]\text{,}
\end{equation}
where one has supervectors and a supermatrix. Note that we have set $l=1$ in our action, which we will continue to do throughout the rest of the paper.
Performing the Gaussian integral yields the partition function
\begin{equation}
Z \propto \left( Sdet
\left[
\begin{array}{c|c}
\ri \slashed{\nabla} & - \sqrt{\frac{2}{N}}  \ri \tilde{\lambda} \\ \hline
\sqrt{\frac{2}{N}}  \ri \lambda &- \partial^2 + 3/4
\end{array}
\right] \right)^{-N}
\text{.}
\end{equation}
A convenient property of superdeterminants is that
\begin{align}
\label{sdetprop}
Sdet
\left[
\begin{array}{c|c}
A & B \\ \hline
C & D
\end{array}
\right]
&=det(A-BD^{-1}C)det(D)^{-1}\text{.}
\end{align}
Using this, the $det(-\partial^2 + 3/4)$ provides an uninteresting overall factor and we are left with computing
\begin{equation}
Z \propto \left[ det\Big(\ri \slashed{\nabla} - \frac{\frac{2}{N} \tilde{\lambda} \lambda}{-\partial^2 + 3/4}\Big) \right]^{-N} \text{.}
\end{equation}
Let us calculate the eigenvalues of 
\begin{equation}
\ri \slashed{\nabla} - \frac{\frac{2}{N} \tilde{\lambda} \lambda}{-\partial^2 + 3/4} \,.
\end{equation}
We are looking for eigenspinors $\chi_n$ with eigenvalues $l_n$ such that

\begin{equation}
\left( \ri \slashed{\nabla} - \frac{\frac{2}{N} \tilde{\lambda} \lambda}{-\partial^2 + 3/4} \right) \chi_n = l_n \chi_n \,.
\end{equation}
We could write $\chi_n$ in a basis of eigenspinors of the Dirac operator. Instead, let us assume that $\chi_n$ is simply an eigenspinor of the Dirac operator with eigenvalue $\Lambda_n$ and we will see that this is the correct solution. Note that eigenspinors of the Dirac operator are also eigenspinors of $-\partial^2 + 3/4$ with eigenvalue $(\Lambda_n)^2$.
  One then has
\begin{equation}
(\Lambda_n)^3 \chi_n^\alpha - \frac{2}{N} \tilde{\lambda}^\alpha \lambda_\beta \chi_{ n}^\beta = l_n (\Lambda_n)^2 \chi_n^\alpha \,.
\end{equation}
Acting with $\tilde{\lambda}^\gamma \lambda_\alpha$ and then dividing out $\lambda_\beta \chi_{ n}^\beta$ yields
\begin{equation}
(\Lambda_n)^3 \tilde{\lambda}^\gamma - \frac{2}{N} \tilde{\lambda}^\gamma \lambda_\alpha \tilde{\lambda}^\alpha = l_n (\Lambda_n)^2 \tilde{\lambda}^\gamma \,,
\end{equation}
which is solved by
\begin{equation}
l_n = \frac{(\Lambda_n)^3 - \frac{2}{N} \lambda_\alpha \tilde{\lambda}^\alpha}{(\Lambda_n)^2} = \frac{(\Lambda_n)^3 - \frac{2}{N} \tilde{\lambda} \lambda}{(\Lambda_n)^2} \,.
\end{equation}
Sensibly, this gives $ \Lambda_n$ when the deformation is turned off.

One can compute an  explicit expression for $Z$ by first treating $ \tilde{\lambda} \lambda$ as a scalar and computing the free energy as a sum using zeta function regularization and then exponentiating. This yields an impressive-looking expression 
which when taking into account the supernumber nature of $\tilde{\lambda}\lambda$ reduces to
\begin{equation}
\label{spinordefpartition}
Z = -\frac{\left(\pi ^4 \left(\pi ^2-10\right) \left( \tilde{\lambda}\lambda \right) ^2-15 N\right) e^{\frac{9 N \zeta (3)}{8 \pi ^2}}}{15 N}.
\end{equation}
Using dS/CFT, this is interpreted as the bulk wave function.

The value of $N_F$, which is calculated from \eqref{NFB}, is given by 
\begin{equation}
N_F = \frac{8 \pi ^8 \left(\pi ^2-10\right)^2}{225 N^2+4 \pi ^8 \left(\pi ^2-10\right)^2} \,.
\end{equation}
We see that for very low values of $N$,  fermionic states are highly occupied in the bulk wave function. With increased $N$, $N_F$ quickly goes to zero. Since, $N = (G_N \Lambda)^{-1}$, with $G_N$ Newton's constant and $\Lambda$ the cosmological constant in the bulk, and we are at constant $\Lambda$, the large $N$ limit corresponds to the free limit. Thus, we find that for a free massless spinor field in de Sitter space, there is no pair production in the Hartle-Hawking ground state. The same result was obtained from a bulk computation of the Hartle-Hawking wave function for a free massless fermion in \cite{DEath:1986lxx,DEath:1996ejf}. This result provides a nontrivial confirmation of the agreement between bulk and boundary computations of the Hartle-Hawking state.
Further, $<\lambda \tilde{\lambda} >=0$ as there is no first order term in $ \lambda \tilde{\lambda}$ in the wave function.

%%%%%%%%%%%%%%%%
\section{Interplay between bulk gauginos and bosons}
\label{secinterplay}
%%%%%%%%%%%%%%%%

So far, we have looked at the bulk spin-1/2 field as the only source turned on and we have seen that it behaves sensibly. Now we will look at the impact that the spin-1/2 field has on a few bosonic bulk deformations. Exact expressions for the eigenvalues of the Laplace and Dirac operators can still be obtained in this case. The complexity of the expressions for these grows and in some cases we were unable to analytically compute the partition function and relied on numerics instead. A subtlety here is that since we need to isolate each order in $\tilde{\lambda}\lambda$ of our result, we cannot rely on numerics so long as there still is $\tilde{\lambda}\lambda$ dependence present in our expressions.

Consider first the case $N=1$. Generically, we will have eigenvalues $\alpha_n(\tilde{\lambda}\lambda)$ and $Z= \prod_n \alpha_n[\tilde{\lambda}\lambda]$. First, we rewrite the eigenvalues as $\alpha_n[\tilde{\lambda}\lambda]=\beta_n[\tilde{\lambda}\lambda]  \gamma_n$, where $\beta_n[0]=1$. We obtain $Z=\prod_n \beta_n[\tilde{\lambda}\lambda] \prod_m \gamma_m$. We define $\beta [\tilde{\lambda}\lambda] \equiv \prod_n \beta_n[\tilde{\lambda}\lambda]$ and $\gamma \equiv \prod_m \gamma_m$ such that $Z = \beta [\tilde{\lambda}\lambda] \gamma$. We now expand $\beta_n$ in powers of $\tilde{\lambda}\lambda$ as $\beta_n = 1 + a_n \tilde{\lambda}\lambda + b_n (\tilde{\lambda}\lambda)^2$.
Now note that 
\begin{equation}
\beta = \prod_n \beta_n = 1 + \left(\sum_n a_n\right)\tilde{\lambda}\lambda + \sum_n \left(b_n + a_n \sum_m a_m\right)(\tilde{\lambda}\lambda)^2\text{.}
\end{equation}
Within the approximations made, the properties of Grassmann numbers have turned an infinite  product into an infinite sum, which is easier to evaluate. These sums and the $\prod_n \gamma_n$ can now be evaluated numerically for a given configuration of the bosonic bulk fields. In what follows we will do this concretely for a few examples

Now, let us use the previous result to obtain a result for arbitrary $N$ in such a way that we will clearly see the $N$ dependence. Firsly, we should remember to divide all the sources by factors of $\sqrt{N}$ as discussed in \cref{sec:deform}. Crucially, this means that all factors of $\tilde{\lambda}\lambda$ will be divided by $N$. Suppose we followed the $N=1$ analysis to compute the partition function $Z[\text{sources}]$ for the $U(-1)$ model. The partition function of the $U(-N)$ model is then given by $Z[\text{sources}/\sqrt{N}]^N$. The overall factor $\prod_m \gamma_m$ in front of every order of $\tilde{\lambda}\lambda$ simply becomes $(\prod_m \gamma_m)^N$. However, when considering the $\beta_n$ sector, we should keep the properties of the Grassmann variables in mind. We see that we get\footnote{Although we have supressed this dependence, remember that $a_n$, $b_n$ and $\gamma_n$ still depend on the bosonic sources, as well as that this dependence will pick up factors of $1/\sqrt{N}$.}
\begin{equation}
\label{betaN}
\ (\prod_n \beta_n)^N = 1 +  \left(\sum_n a_n\right)\tilde{\lambda}\lambda + \left( \frac{1}{N}\sum_n \left(b_n + a_n \sum_m^{n-1} a_m\right) +\frac{ (N^2-N)}{N^2}\frac{ \left(\sum a_n\right)^2}{2}\right) (\tilde{\lambda}\lambda)^2 \text{.}
\end{equation}
We see that there is an interesting transition in the behaviour of the $(\tilde{\lambda}\lambda)^2$ term between small and large $N$. Note that if $\sum_n a_n = 0$, the gaugino contribution to the wave function is supressed at large $N$. 
This is precisely what happened when we considered turning on only the gaugino deformation without any other deformations.

%%%%%%%%%%%%%%%%
\subsection{Bulk pseudoscalar}
\label{pseudoscalarsec}
%%%%%%%%%%%%%%%%

The simplest deformation to combine with turning on the gauginos consists of giving a mass to the $\chi$ spinors. This corresponds to turning on the pseudoscalar field in the bulk. The action for $N=1$ is then of the form
\begin{equation}
S = \int d^3 x \sqrt{h}
\left[ 
\begin{array}{c|c}
\tilde{\chi} &  \tilde{\varphi}
\end{array}
\right]
\left[
\begin{array}{c|c}
\ri \slashed{\nabla} + m & - \sqrt{2}  \ri \tilde{\lambda} \\ \hline
\sqrt{2}  \ri \lambda &- \partial^2 + 3/4
\end{array}
\right]
\left[
\begin{array}{c}
\chi \\ \hline
\varphi
\end{array}
\right]\text{.}
\end{equation}

Proceeding in the same way as the computation with only the gauginos, we see that the partition function is of the form
\begin{equation}
Z \propto \left[ det\Big(\ri \slashed{\nabla} + m - \frac{\lambda \tilde{\lambda}}{\partial^2 + 3/4}\Big) \right]^{-1} \text{,}
\end{equation}
which means we need to know the eigenvalues of
\begin{equation}
\ri \slashed{\nabla} + m - \frac{\tilde{\lambda}\lambda }{\partial^2 + 3/4}
~.
\end{equation}
By the same procedure as used for just the gauginos in \cref{Spin1/2}, we find for these eigenvalues 
\begin{equation}
l_n= \frac{(\Lambda_n)^3 + m (\Lambda_n)^2 -2\tilde{\lambda}\lambda}{(\Lambda_n)^2}\,.
\end{equation}
Then, $Z$ is given by
\bea
Z\propto \prod_n a_n^{-1} =\b\, \g
~,~~~~~~
\b:= \prod_n \frac{(\Lambda_n)^3 + m (\Lambda_n)^2}{(\Lambda_n)^3 + m (\Lambda_n)^2 - 2  \tilde{\lambda \lambda}} 
~,~~~
\g:=\prod_n \frac{(\Lambda_n)^2}{(\Lambda_n)^3 + m \, (\Lambda_n)^2}
~.
\eea

We evaluated $\b$ numerically and obtained the following analytic expression for $\g$
\begin{equation}
\gamma \propto \sqrt[4]{\cos (\pi  m )} \exp \left(\frac{\ri \pi  m  \text{Li}_2\left(-e^{2 \ri \pi  m }\right)
-\frac{1}{2} \text{Li}_3\left(-e^{2 \ri \pi  m }\right)+\frac{1}{3} \ri \pi ^3 m ^3-\pi ^2 m^2 \log \left(1+e^{2 \ri \pi  m }\right)}{\pi ^2}\right) 
\,.
\end{equation}

\begin{figure}
\includegraphics[scale=0.6]{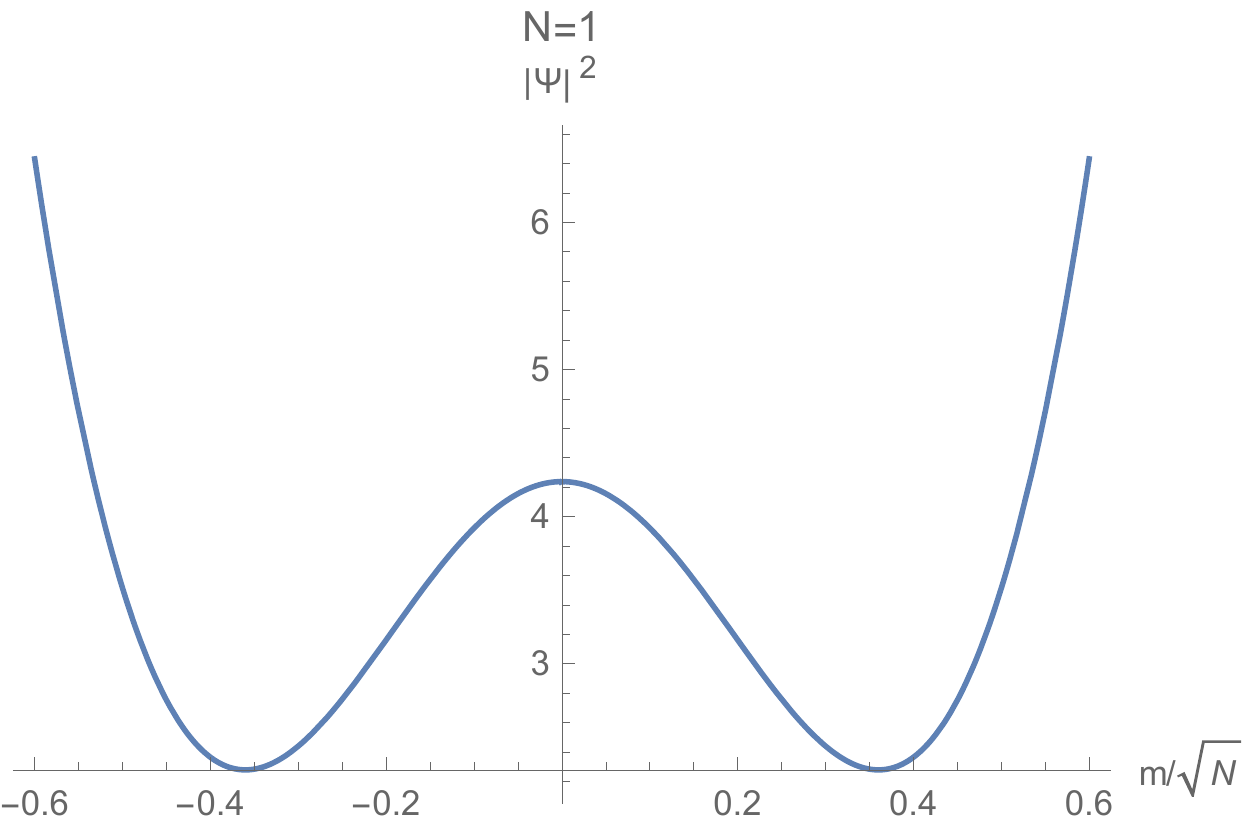}
\includegraphics[scale=0.6]{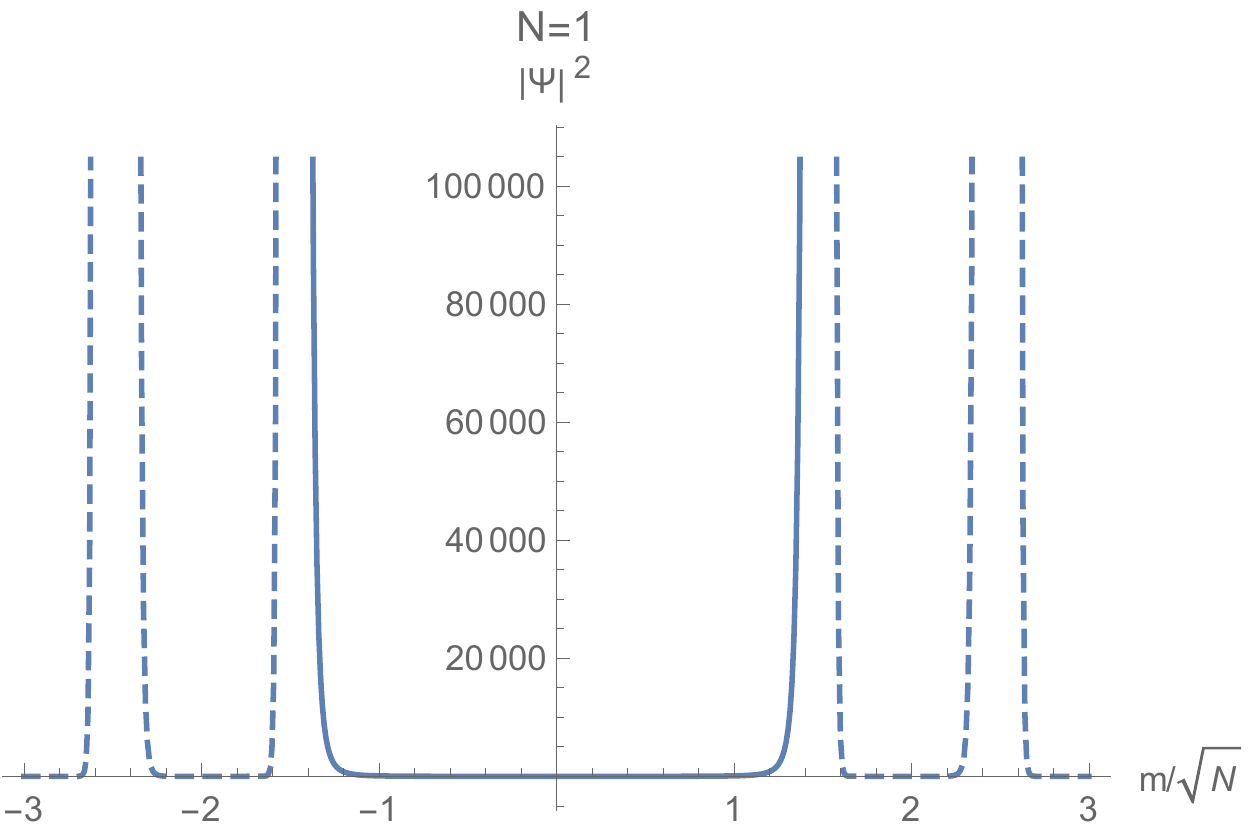}

\caption{For a deformation corresponding to turning on a bulk pseudoscalar field, $|\Psi^2 (m)|$ over a short and a long range at $N=1$. The point $m=0$ has been excluded from the short range plot due to problems with the numerical evaluation of very small numbers.}
\label{pseudoscalarwavshort}
\end{figure}

\begin{figure}
\includegraphics[scale=0.6]{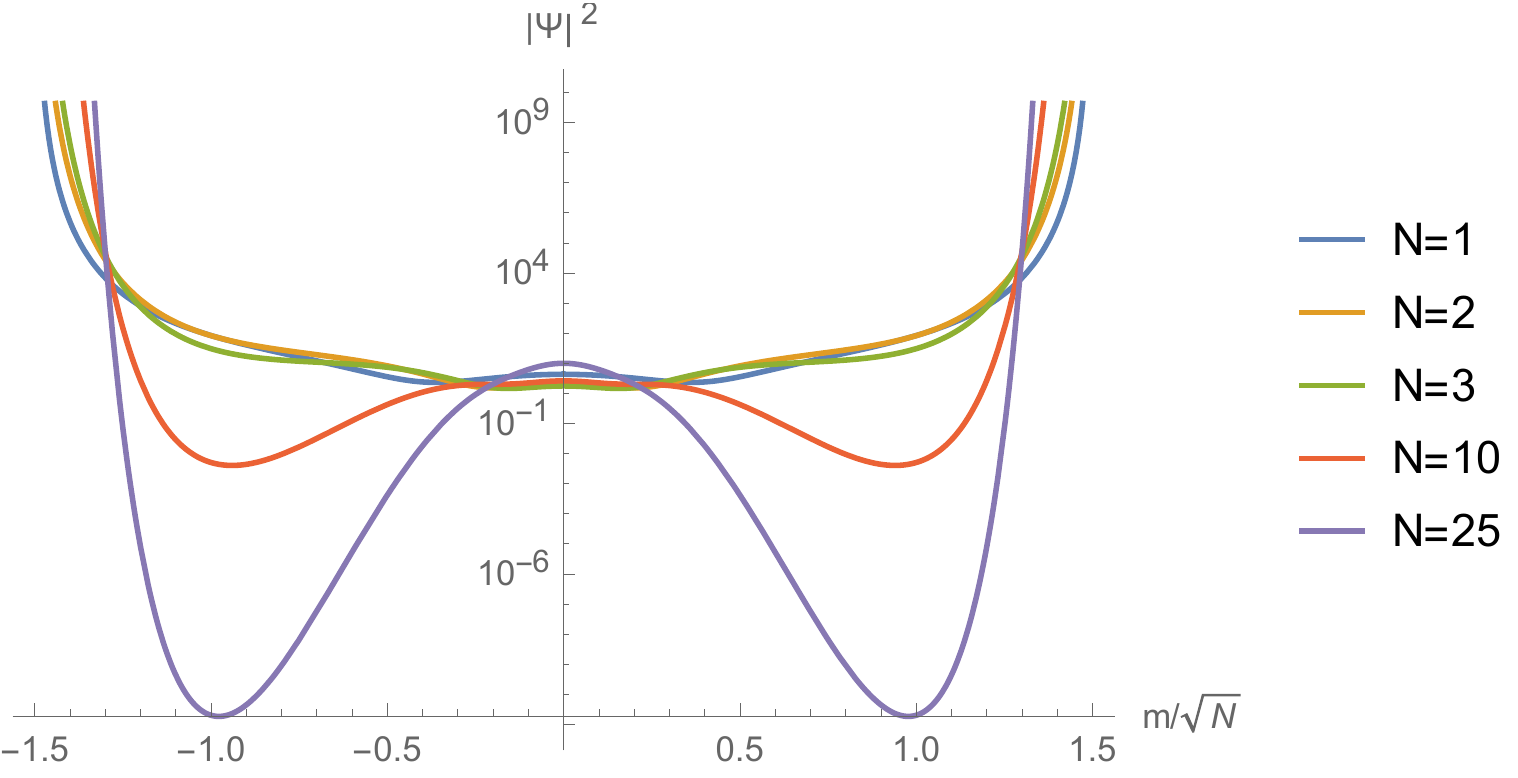}
\caption{For a deformation corresponding to turning on a bulk pseudoscalar field, $|\Psi^2 (m)|$.}
\label{pseudoscalarwavlong}
\end{figure}

\begin{figure}
\includegraphics[scale=0.6]{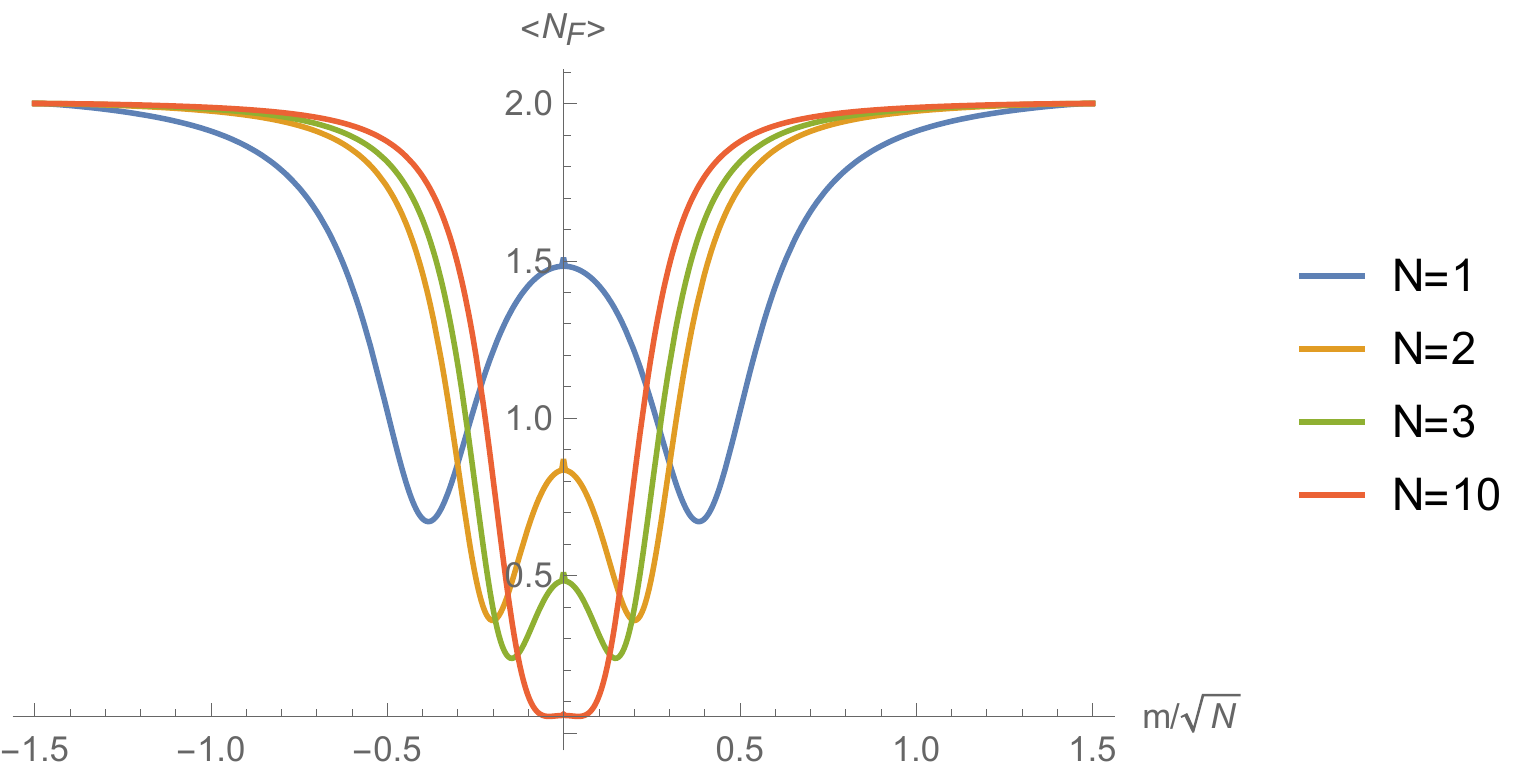}
\caption{For a deformation corresponding to turning on a bulk pseudoscalar field, $N_F(m)$.}
\label{pseudoscalarnum}
\end{figure}

\begin{figure}
\includegraphics[scale=0.5]{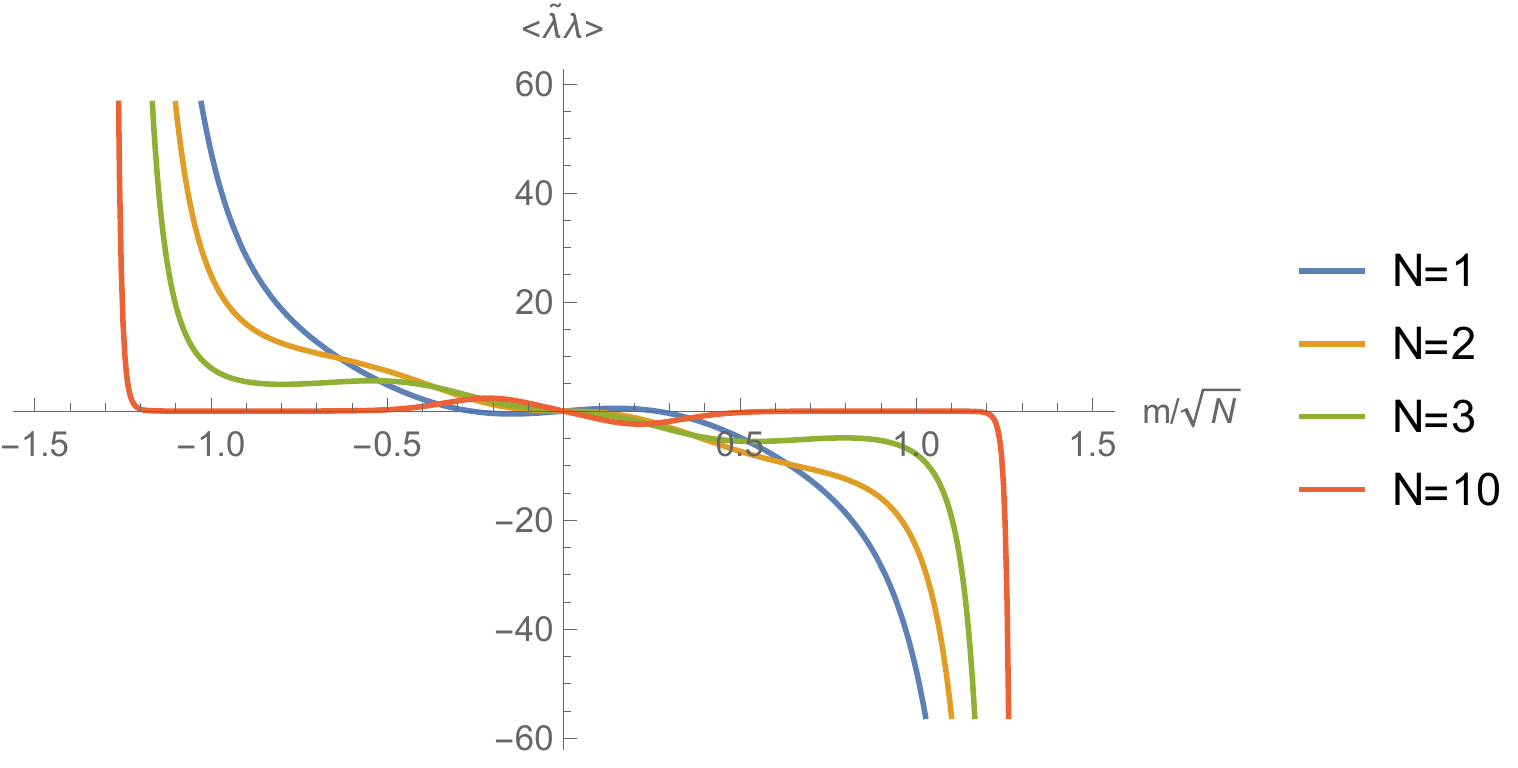}
\includegraphics[scale=0.5]{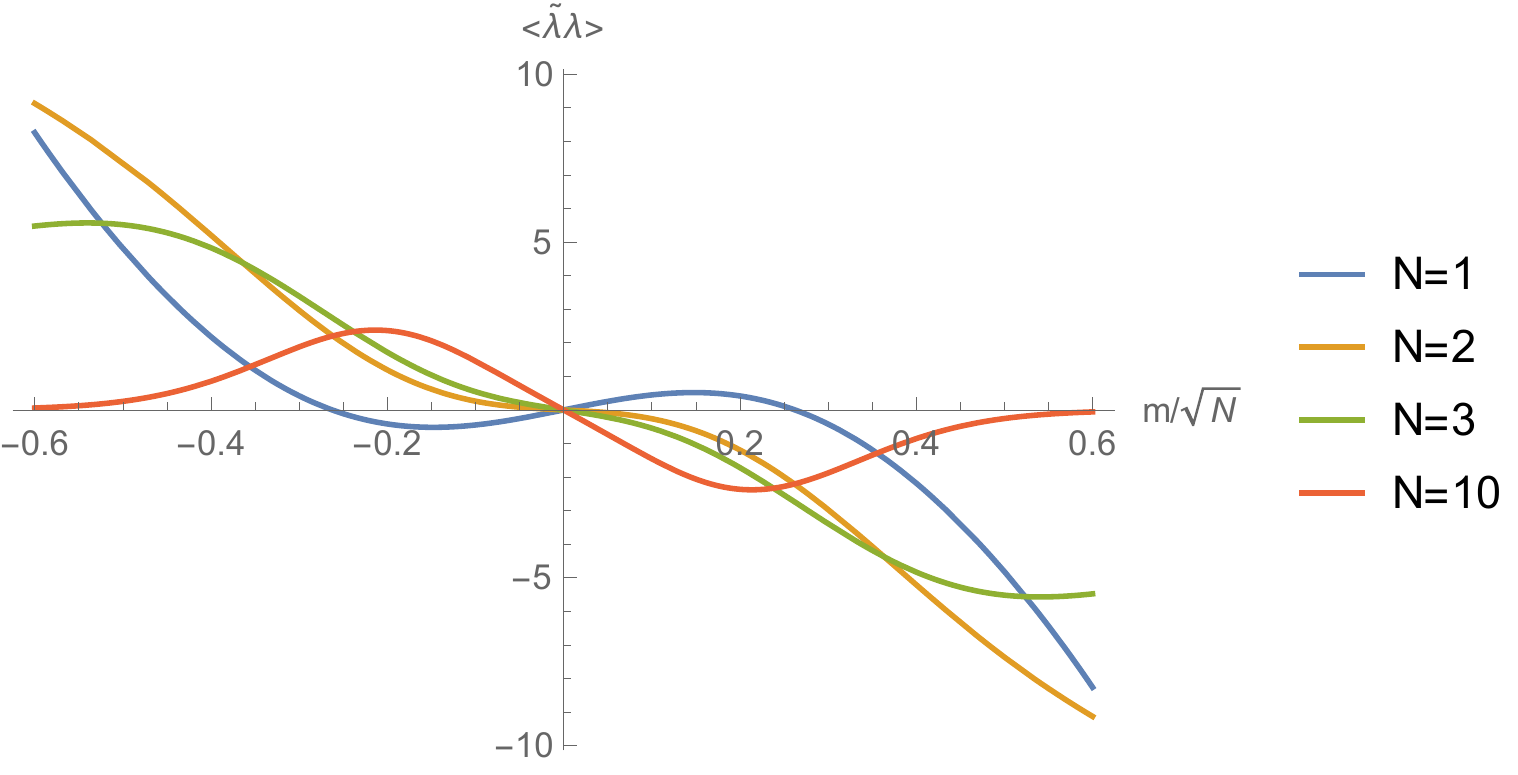}
\caption{For a deformation corresponding to turning on a bulk pseudoscalar field, $<\tilde{\lambda} \lambda>[m]$.}
\label{pseudovevlong1}
\end{figure}

The behaviour of $|\Psi |^2$ can be seen in \cref{pseudoscalarwavshort,pseudoscalarwavlong}. We see that undeformed de Sitter space is a local maximum of $|\Psi |^2$, indicating the perturbative stability of the supersymmetric de Sitter vacuum. For very large deformations, we see divergences in $|\Psi |^2$. These divergences do not come as a surprise. They are not a result of the inclusion of the gauginos, they are already present when one considers the bulk pseudoscalar field by itself, as we already discussed in \cite{Hertog:2017ymy}. In fact, we should restrict the outermost edge of the configuration space at which our results can be hoped to be trusted to $|m/\sqrt{N}| < 3/2$ and we restrict our analysis of observables to this range. We defer the discussion of the reason for this restriction to \cref{narrowdiscussion}

In \cref{pseudoscalarnum} the value of $<N_F>[m]$ can be seen.  We shall interpret its behaviour and compare it to earlier bulk results in \cref{narrowdiscussion}.

In \cref{pseudovevlong1}, $<\tilde{\lambda} \lambda>[m]$ can be seen. We see that it increases as we increase $|m|$. We also see that for $m/\sqrt{N}$ between $-3/2$ and $3/2$, it is antisymmetric, while $|\Psi |^2$ is symmetric in $m$. While we do not know the exact measure against which we should integrate the bosonic fields in the wave function, it seems reasonable that it should respect such symmetries. If this is the case, $<\tilde{\lambda} \lambda>=0$ for the holographic no-boundary wave function after integration over $m$. 
%%%%%%%%%%%%%%%%
\subsection{Bulk scalar}
\label{scalarsec}
%%%%%%%%%%%%%%%%
 
 For a deformation based on turning on a bulk scalar, 
 the action at $N=1$ takes the form

\begin{equation}
\label{gauginomatrixaction-2}
S = \int d^3 x \sqrt{h}
\left[ 
\begin{array}{c|c}
\tilde{\chi} &  \tilde{\varphi}
\end{array}
\right]
\left[
\begin{array}{c|c}
\ri \slashed{\nabla} & - \sqrt{2}  \ri \tilde{\lambda} \\ \hline
\sqrt{2}  \ri \lambda &- \partial^2 + 3/4 + \sigma
\end{array}
\right]
\left[
\begin{array}{c}
\chi \\ \hline
\varphi
\end{array}
\right]\text{,}
\end{equation}
To compute the partition function 
\bea
Z\propto \prod_n a_n^{-1} =\b\, \g
~,~~~
\b:= \prod_n \frac{(\Lambda_n)^3 + \s \Lambda_n}{(\Lambda_n)^3 + \s \Lambda_n - 2  \tilde{\lambda \lambda}} 
~,~~~
\g:=\text{det}(- \partial^2 + 3/4 + \sigma)\prod_n\frac{1}{(\Lambda_n)}
~,~~~~~~
\eea
we proceed in the same way as the previous cases. Note that in this case, the $det(D)^{-1}$ term in \cref{sdetprop} is not constant, but depends on $\sigma$, so we must take it into account. We find for $N=1$
\begin{align}
\beta
&= 1 
-\frac{\pi  \text{sech}^2\left(\pi  \sqrt{\sigma}\right) \left(8 \pi  \sigma^{3/2}
+4 \pi  \sqrt{\sigma}
-(4 \sigma +3) \sinh \left(2 \pi  \sqrt{\sigma}\right)
+2 \pi  \sqrt{\sigma} \cosh \left(2 \pi  \sqrt{\sigma}\right)\right)}{32 \sigma^{5/2}} (2\tilde{\lambda}\lambda)^2 \,,
\end{align}
and
\begin{align}
\gamma \propto \exp \Bigg\{-\frac{1}{24 \pi ^2}\bigg{[}&\,
\pi ^2 \left(\ri \pi  \left(4 \sigma \sqrt{1-4 \sigma}-\sqrt{1-4 \sigma}+1\right)+6 (4 \sigma-1) \log \left(1-e^{-\ri \pi  \sqrt{1-4 \sigma}}\right)\right)
\non\\
&
-12 \ri \pi  \sqrt{1-4 \sigma} \text{Li}_2\left(e^{-\ri \sqrt{1-4 \sigma} \pi }\right)-12 \text{Li}_3\left(e^{-\ri \sqrt{1-4 \sigma} \pi }\right)
\bigg{]}\Bigg\} \,.
\end{align}
We again expand $\beta$ via the methods described previously. In this case, it was possible to find $\beta$ analytically. It is worth noting that $\sum_n a_n = 0$. Looking at \cref{betaN}, this ensures that in the large $N$ limit, the parts of the wave function with bulk fermions excited are supressed and the local fermion excitation number tends to zero. Of course, any actual divergences introduced by the gauginos remain at any finite $N$.

\begin{figure}
\includegraphics[scale=0.6]{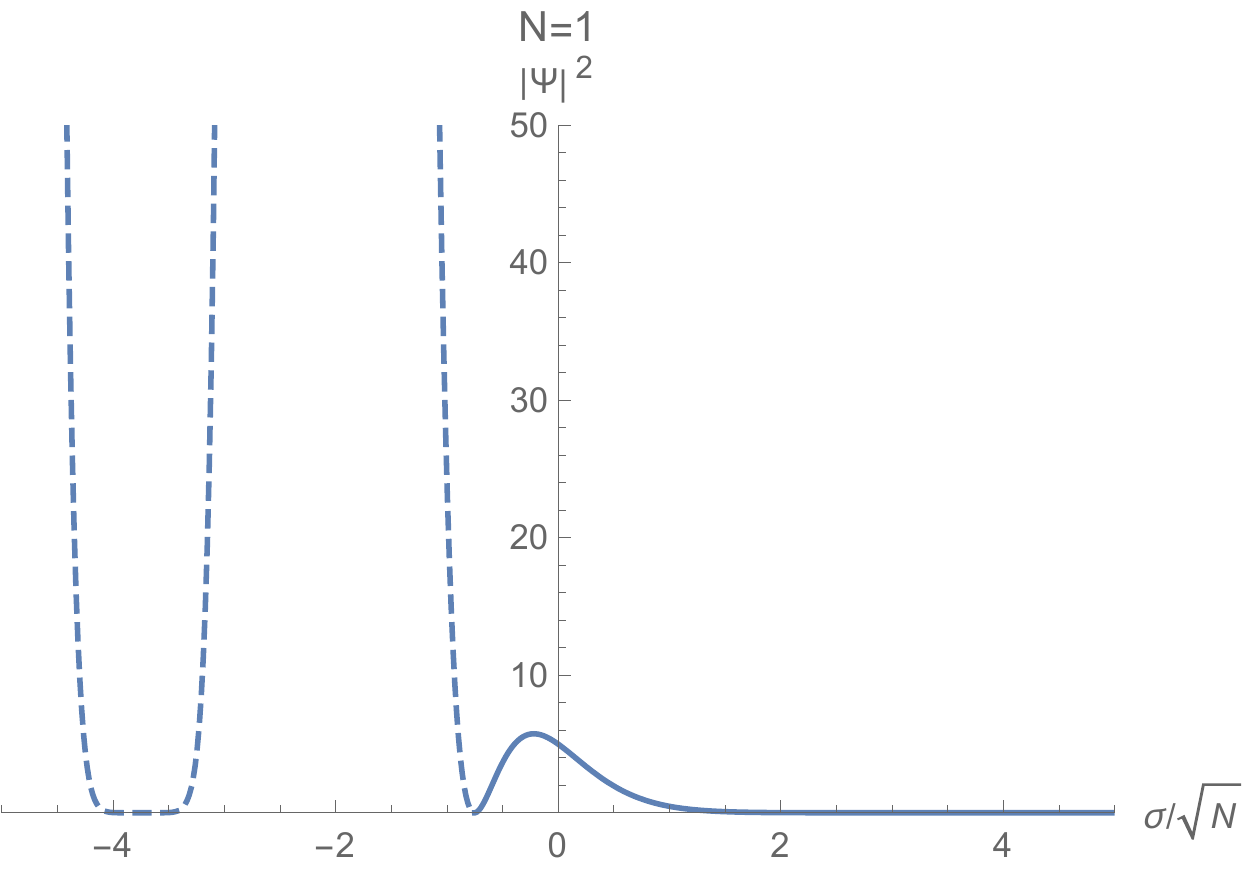}
\includegraphics[scale=0.6]{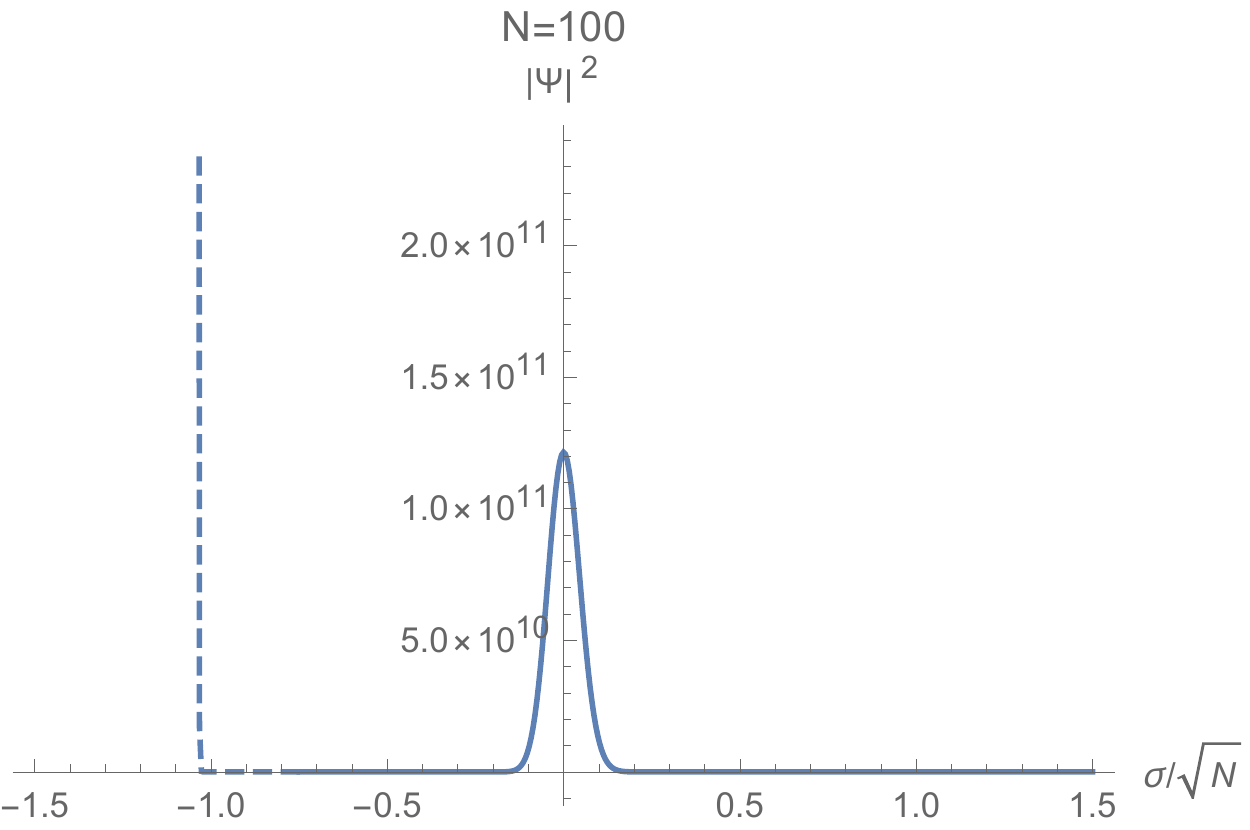}
\includegraphics[scale=0.6]{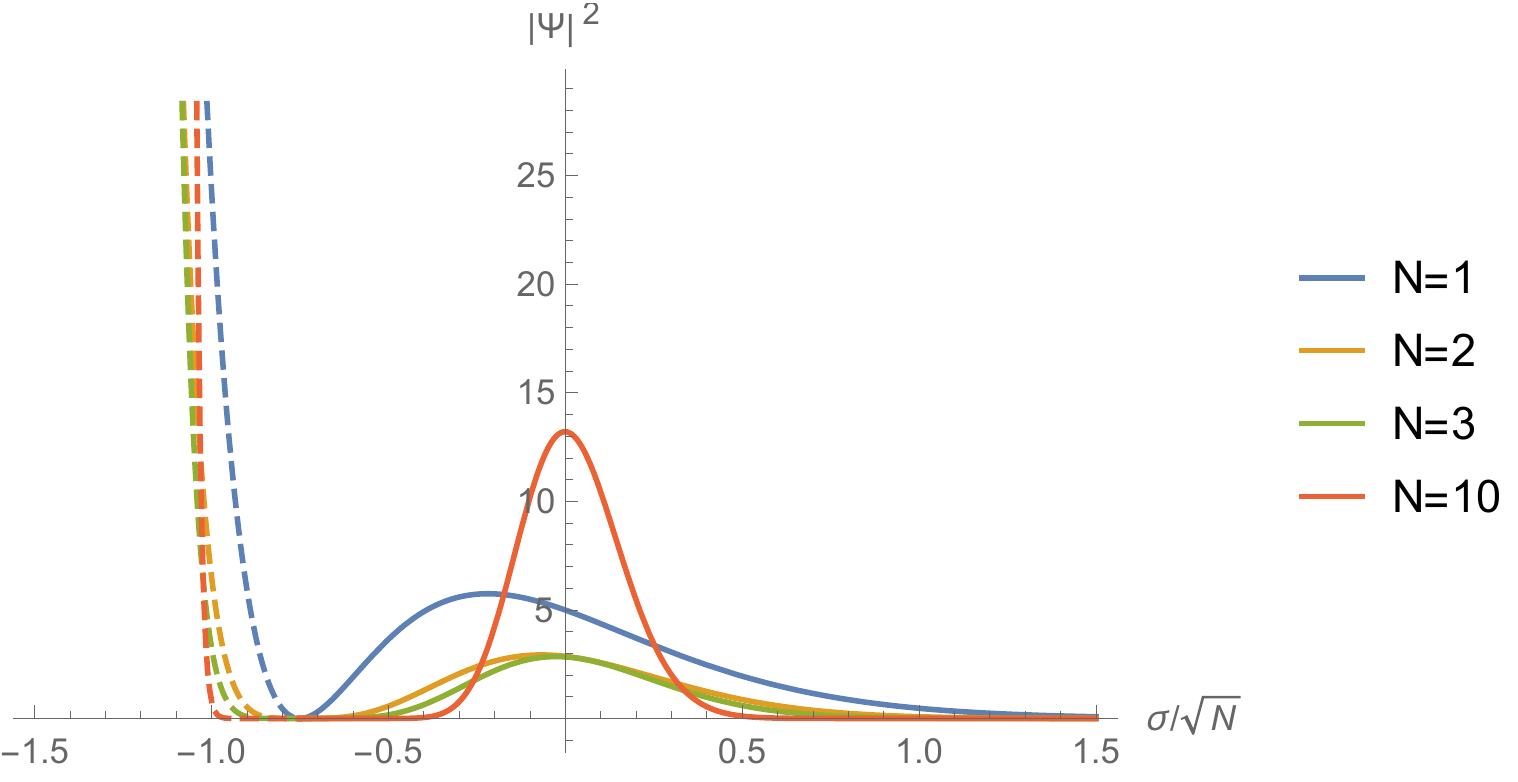}
\caption{For a deformation corresponding to turning on a bulk scalar field, $|\Psi^2 (\sigma )|$.}
\label{scalarwav}
\end{figure}

Without the gauginos but with only the scalar mass deformation, our deformation would be essentially equivalent to the scalar deformation considered at the start of \cite{Anninos:2012ft}. As such, it makes sense to compare this wave function to the one discussed there. For $N=1$, $|\Psi |^2$ can be seen in \cref{scalarwav}. We see that for small deformation, there is a local maximum for $|\Psi |^2$ at undeformed de Sitter space. For positive deformations, this maximum is global. For negative deformations, we see large fluctuations, as were also found in \cite{Anninos:2012ft}. Here the problem seems to be exacerbated. Rather than large finite fluctuations, $|\Psi |^2$ now 
only diverges for sufficiently large negative deformations, with the first divergency at $\sigma/\sqrt{N}=-9/4$. As we shall discuss in \cref{narrowdiscussion}, we should restrict the configuration space to $\sigma/\sqrt{N}> -3/4$.

\begin{figure}
\includegraphics[scale=0.6]{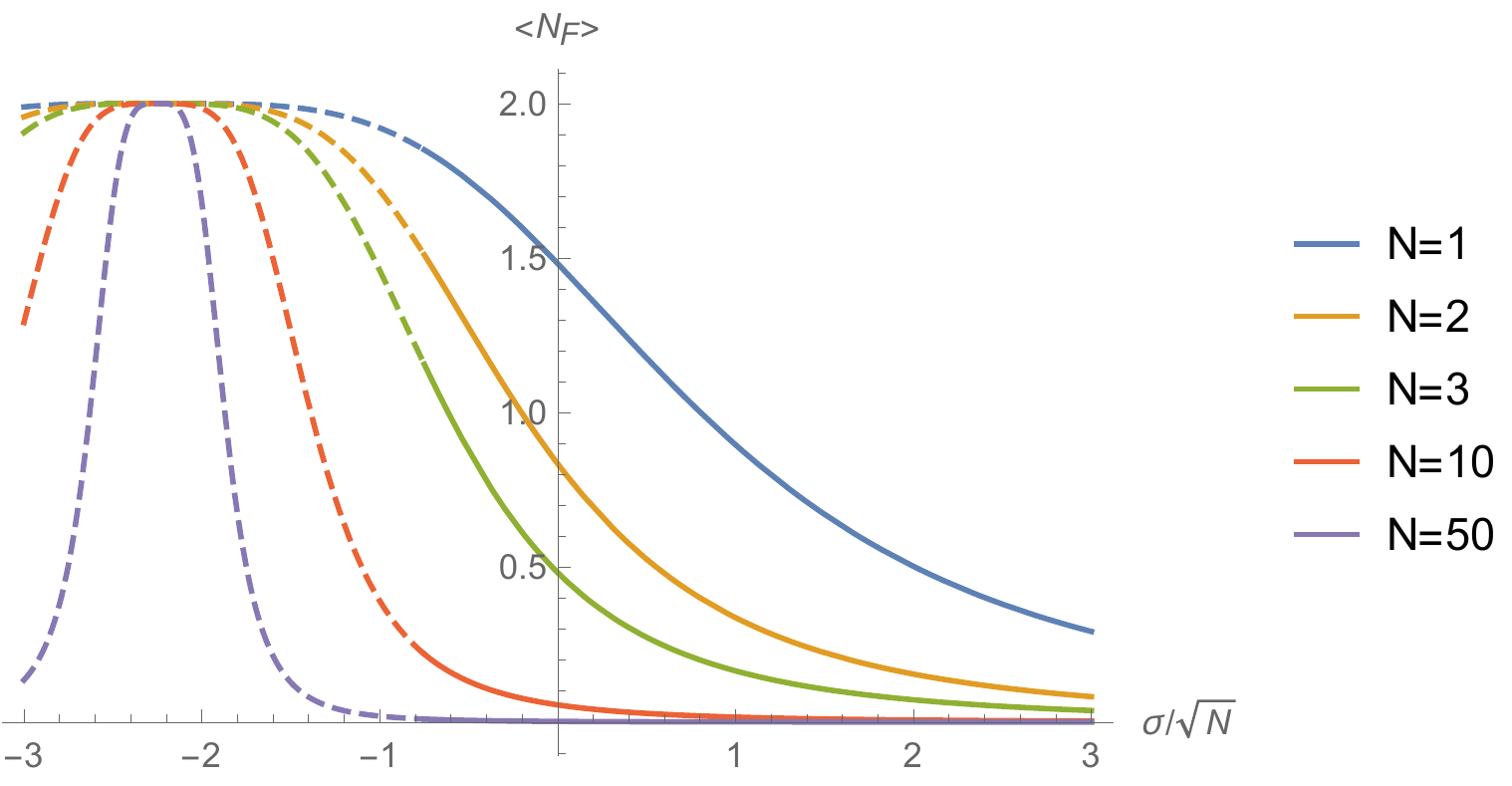}
\caption{For a deformation corresponding to turning on a bulk scalar field, $N_F(\sigma )$.}
\label{scalarnum}
\end{figure}

The fermion number can be seen in \cref{scalarnum}, we postpone its analysis to  \cref{narrowdiscussion}.  We see that $<\tilde{\lambda} \lambda>=0$ as there is no first order term in $\tilde{\lambda} \lambda$ in the wave function.

%%%%%%%%%%%%%%%%
\subsection{Squashing}
\label{squashingsec}
%%%%%%%%%%%%%%%%

Finally, we combine the spin-1/2 deformation with a squashing of the three-sphere. The action takes the form
\begin{equation}
S = \int d^3 x \sqrt{h_{\text{def}}}
\left[ 
\begin{array}{c|c}
\tilde{\chi} &  \tilde{\varphi}
\end{array}
\right]
\left[
\begin{array}{c|c}
\ri \slashed{\nabla} & - \sqrt{\frac{2}{N}}  \ri \tilde{\lambda} \\ \hline
\sqrt{\frac{2}{N}}  \ri \lambda &- \partial^2 + \frac{R}{8}
\end{array}
\right]
\left[
\begin{array}{c}
\chi \\ \hline
\varphi
\end{array}
\right]\text{.}
\end{equation}
Turning on the squashing changes the eigenvalues of the Dirac operator and Laplacian. We consider a squashing preserving $SU(2) \times U(1)$ symmetry described by the metric
\begin{equation}
ds_{S^{3}_{\alpha}}^2 = d\theta^2+\sin^2\theta d\phi^2\,+\frac{1}{1+\alpha}(d\tilde{\psi}+\cos \theta d\phi)^2 \, , 
\end{equation}
where  $\theta \in [0,\pi]$, $\phi \in [0,2\pi]$, 
$\tilde{\psi} \in [0,4\pi]$, and the real deformation parameter $\alpha$ is such that $\alpha>-1$.

The eigenvalues of the Dirac operator are of the form \cite{Bobev:2017asb, Dowker:1998pi, Gibbons:1979kq, HITCHIN19741}
\begin{equation}
\Lambda_{n,q,\,\pm} =   \frac{1}{2 \sqrt{1+\alpha}}  \pm 2\sqrt{ \frac{n^2(1+\alpha)}{4}-\alpha q (n - q) }  \, ,
\end{equation}
with degeneracy $n$. Here the eigenvalues run over two branches. For the positive branch, $n=1,..,\infty$ and $q=0,..,n$ while for the negative branch, $n=2,..,\infty$ and $q=1,..,n-1$.
To compute the partition function, the same logic as in the previous sections applies, except that we use the eigenvalues of the Dirac operator on the squashed three-sphere. We find for $N=1$
\begin{equation}
\beta = \prod_{n,q} \frac{\Lambda_{n,q}^3}{\Lambda_{n,q}^3- 2 \tilde{\lambda}\lambda}\,.
\end{equation}
The gaugino independent part of the partition function, $\gamma$, is the partition function for a free spinor and complex scalar on a squashed three sphere. This was computed numerically in \cite{Bobev:2017asb}
where a good agreement  was found between the numerical result and the following analytic expression 
which we will use for $\gamma$,
\begin{equation}
\gamma = -(-1)^{23/24} \sqrt[4]{2} e^{\frac{3 \zeta (3)}{4 \pi ^2}-\frac{\pi ^2 \alpha ^2}{16 (\alpha +1)^2}} \,.
\end{equation}
We compute $\beta$ numerically, analogously to the previous sections, in this way obtain $Z = \beta \gamma$ and identify it with the wave function under dS/CFT.

\begin{figure}
\includegraphics[scale=0.6]{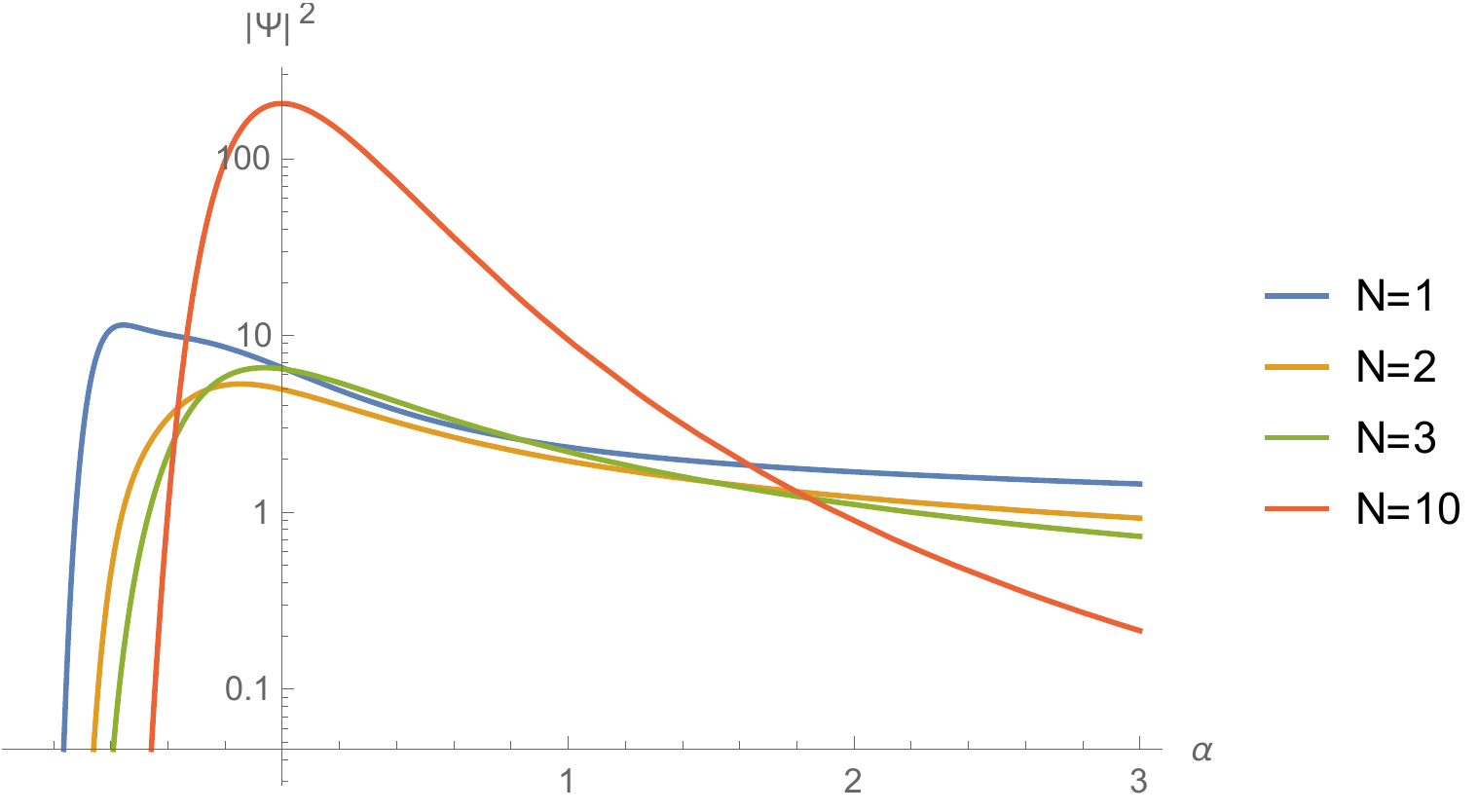} \includegraphics[scale=0.6]{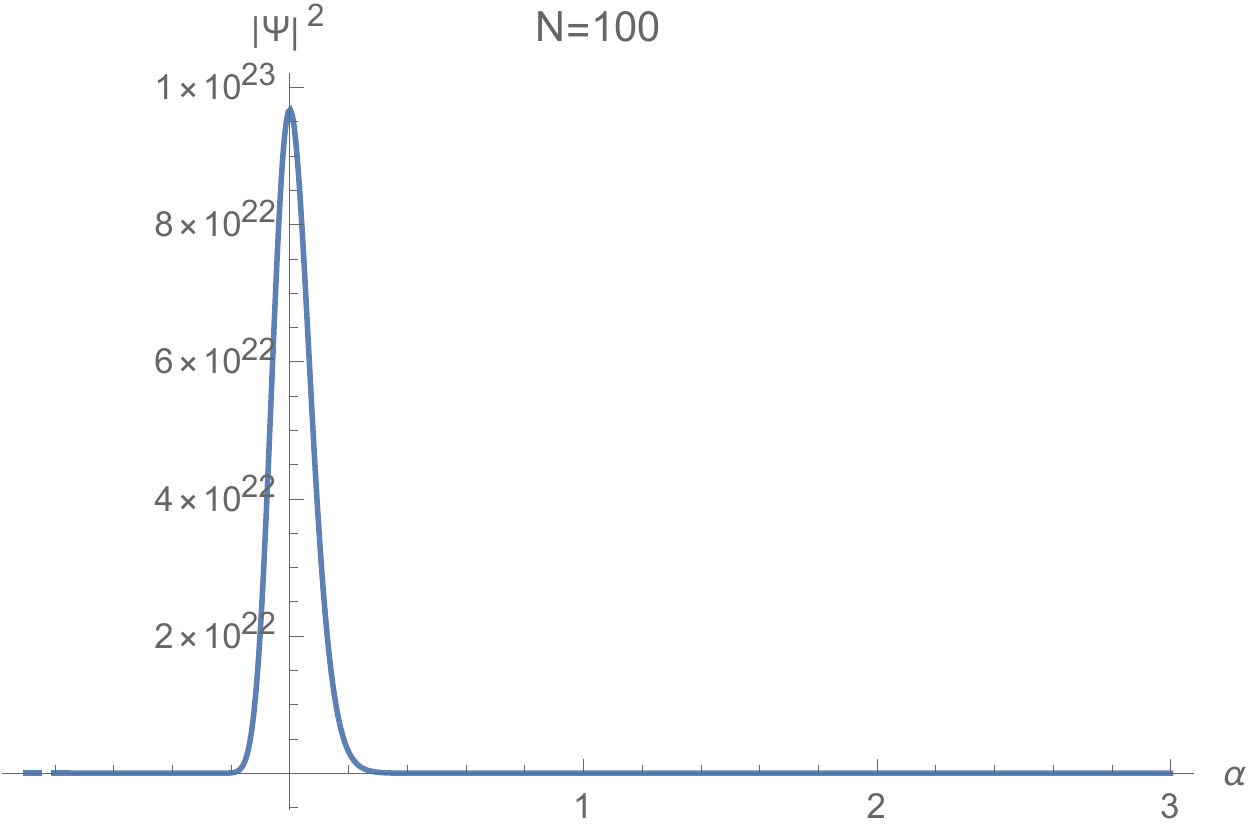}
\caption{For a squashing, $|\Psi (\alpha )|^2$ .}
\label{squashwav}
\end{figure}

The behaviour of $|\Psi |^2$ under squashing can be seen in  \cref{squashwav}. We find an interesting dependence on $N$. At small $N$, $|\Psi |^2$ is peaked for a squashed three-sphere. As $N$ increases, we see that the peak of $|\Psi |^2$ quickly moves back to te undeformed three-sphere. At large\footnote{We see  that at, for instance, $N=10$, large $N$ effects are totally dominant. The traditional wisdom that ``three is a large number'' when it comes to large $N$ expansion seems to apply here too.} $N$, $|\Psi |^2$ has a global maximum at undeformed de Sitter space, indicating the stability of the de Sitter vacuum. At small $N$, the stable vacuum around which one should consider perturbation theory seems to be squashed away from the three-sphere. For reasons discussed in  \cref{narrowdiscussion}, we should restrict the configuration space to $\alpha > -3/4$

\begin{figure}
\includegraphics[scale=0.6]{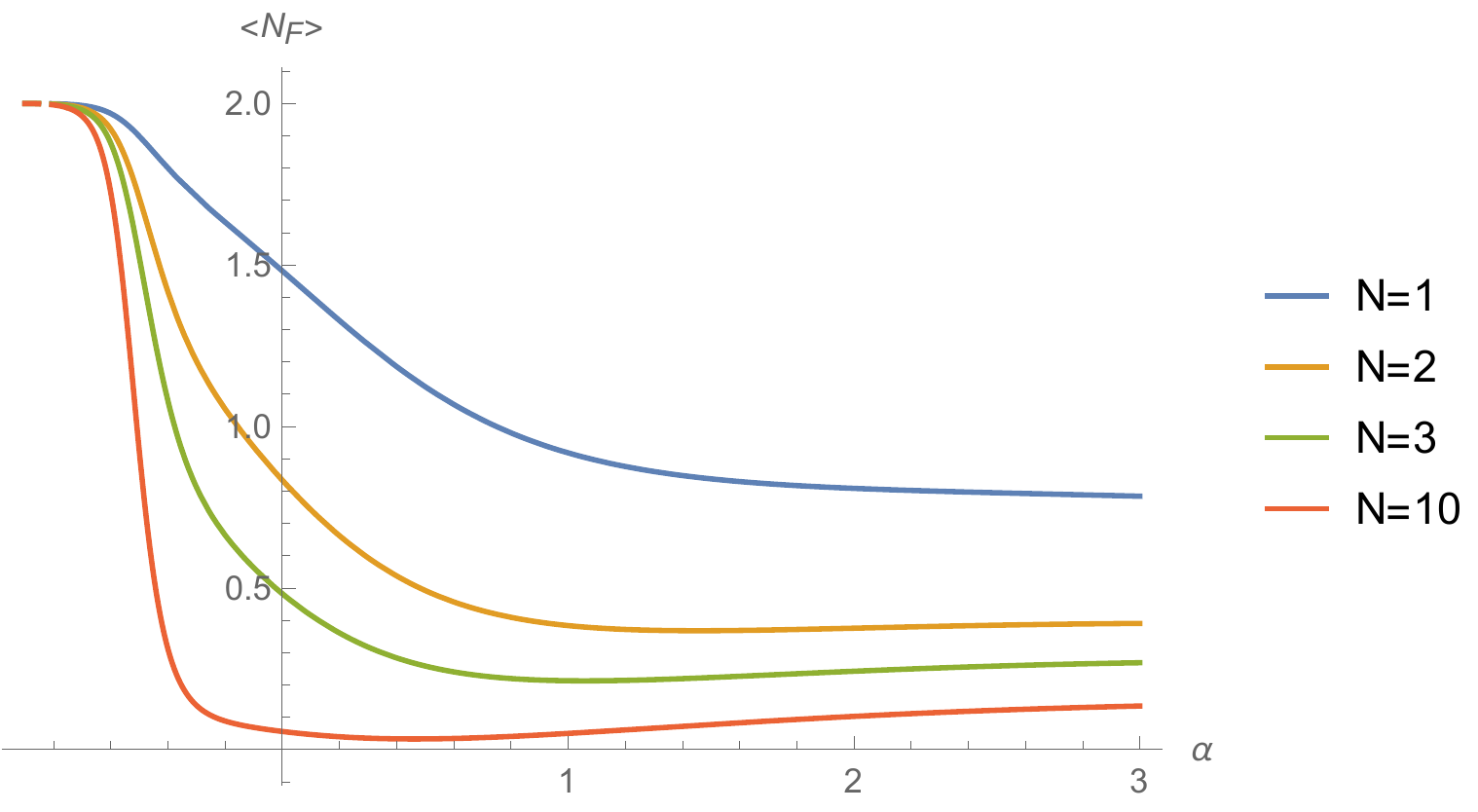}
\caption{For a squashing, $N_F(\alpha )$.}
\label{squashNF}
\end{figure}

\begin{figure}
\includegraphics[scale=0.6]{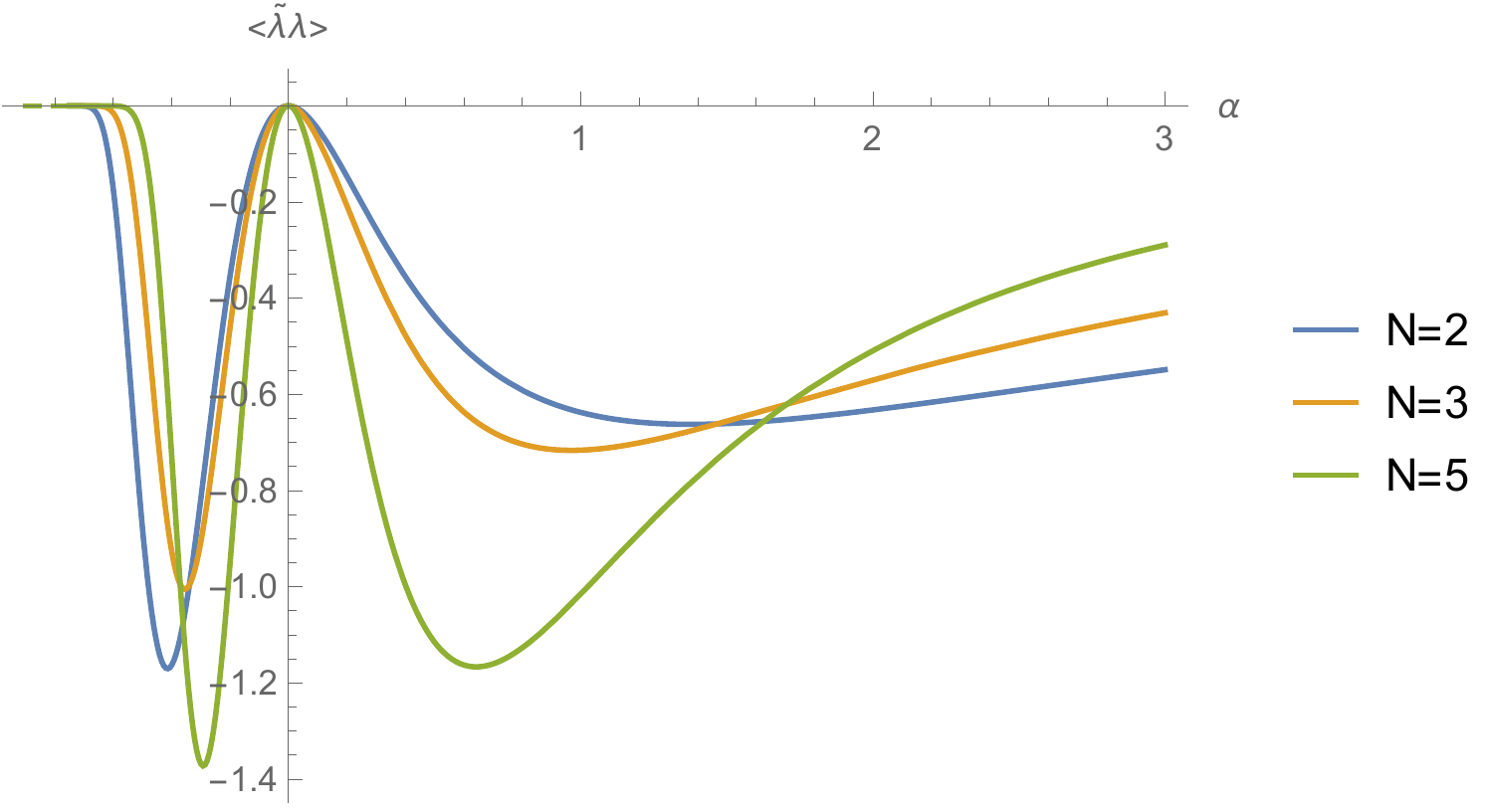}
\caption{For a squashing, $<\tilde{\lambda} \lambda>$ .}
\label{squshingvev_N2_3_5}
\end{figure}

The fermion occupation number, whose properties we will discuss further in  \cref{narrowdiscussion}, can be seen in \cref{squashNF}. In \cref{squshingvev_N2_3_5}, $<\tilde{\lambda} \lambda>[B]$ can be seen. We see that it is negative and reaches an extremum at finite squashing.

%%%%%%%%%%%%%%%%
\subsection{Interpretation and Comparison}
\label{narrowdiscussion}
%%%%%%%%%%%%%%%%

\begin{table}[]
\begin{tabular}{|l|l|l|}
\hline
\multicolumn{1}{|l|}{\textbf{Deformation}} & \textbf{Effect on wavefunction} & \multicolumn{1}{l|}{\textbf{Studied in}} \\ \hline
Scalar                        &   Perturbatively stable around empty de Sitter                     & \cite{Anninos:2012ft} \\ \hline
BPS scalar + pseudoscalar     &   Perturbatively stable around empty de Sitter                     & \cite{Hertog:2017ymy} \\ \hline
BPS vector + metric squashing &    Perturbatively stable around empty de Sitter                    & \cite{Hertog:2017ymy} \\ \hline
Spinor                        &  Fermion states occupied at small $N$, unoccupied at large $N$                      & \cref{Spin1/2} \\ \hline
Spinor + pseudoscalar         &Perturbatively stable around empty de Sitter                        & \cref{pseudoscalarsec} \\ \hline
Spinor + scalar               &Perturbatively stable, peak shifted from empty dS at small $N$                        & \cref{scalarsec} \\ \hline
Spinor + metric squashing     & Perturbatively stable, peak shifted from empty dS at small $N$                        & \cref{squashingsec}\\ \hline
\end{tabular}
\caption{Summary of for which bulk fields the wavefunction of the $\mathcal{N}=2$ supersymmetric higher spin theory has been computed using dS/CFT. The pseudoscalar field or squashing by itself in the SUSY model have not been given a detailed discussion, but its behaviour can easily be seen by setting $\beta=1$ in the relevant section.}
\label{summarytable}
\end{table}

We have computed the Hartle - Hawking wave function for an asymptotically nearly de Sitter universe in a `minisuperspace' model consisting of the zero mode of a spin-1/2 field, a pseudoscalar, a scalar and a metric deformation as excitations around dS space. A summary of the deformations studied in this paper and the earlier literature can be seen in \cref{summarytable}.  The wave function predicts probabilities for different asymptotic configurations (or more precisely, different spacetime histories). The \cref{pseudoscalarwavshort,pseudoscalarwavlong,scalarwav,squashwav} show that the probability distributions exhibit a local maximum for reasonably small deformations in all directions of the minisuperspace. At small $N$, this maximum can be shifted somewhat away from the undeformed dS configuration. At large $N$, however, it always lies perfectly at the empty, undeformed de Sitter space. In this semiclassical limit the wave function becomes increasingly sharply concentrated around this configuration. The presence of a local maximum at empty dS also indicates that dS is perturbatively stable. The behaviour of the wave function for small fluctuations around dS describes correlators of fields in dS, and a maximum ensures these behave correctly.

For integer-spin sources, the fact that the probabilities are peaked at empty undeformed de Sitter space follows basically from an analytic continuation of the F-theorem \cite{Jafferis:2010un,Jafferis:2011zi,Klebanov2011,Casini:2012ei}. This states that for unitary QFTs on $S^3$, the free energy is maximized by a UV CFT and any deformation - including complex deformations - decreases it. This theorem has been proven and extended to a variety of CFT deformations, e.g. the extension in \cite{Bobev:2017asb} is especially relevant for us. With the spin-1/2 source deformation included, it is at small $N$ no longer true that the wave function squared is peaked at empty undeformed de Sitter space. It is also not clear whether the wave function has a meaningful interpretation in terms of probabilities in this limit. In fact, with a spin-1/2 source, the partition function is supernumber-valued and there is no longer an unambiguous notion of maximum for $Z$. First, $Z$ must be mapped to the real numbers to speak of maximization. The inner product \cref{innerproduct} provides such a map, but there is no reason why an F-theorem should imply maximization at the undeformed CFT with respect to this specific inner product. Still, there is a notion of $Z$ being extremized with respect to the spin-1/2 deformation at the undeformed CFT. Clearly the first derivative of \cref{spinordefpartition} with respect to $\lambda$, $\tilde{\lambda}$, or both, is zero at $\lambda=0$ or $\tilde{\lambda}=0$. This implies that $<\tilde{\lambda}\lambda>=0$.

For large scalar or pseudoscalar deformations our results may give the impression that the probabilities exhibit divergences. However, as we already mentioned in \cref{secinterplay}, dS/CFT comes with a significant restriction on the configuration space of deformations which in particular excludes the regime where we find the wave function diverges. These restrictions are indicated with a transition to dotted lines in the figures above.

For the scalar deformation, we can identify the boundaries of the configuration space with a good degree of confidence. Here, the completion of dS/CFT of \cite{Anninos:2017eib} which provides not just a wave function but also a Hilbert space and an inner product, directly applies to our model. It implies a measure that restricts the configuration space of the wave function to deformations with $\sigma / \sqrt{N} > -3/4$.  In \cite{Hertog:2017ymy} we gave an independent argument for the same bound based on the origin of dS/CFT as a continuation of Euclidean AdS/CFT. At $\sigma / \sqrt{N} = -3/4$ one of the eigenvalues of the dynamical CFT scalar becomes zero, and below this it becomes negative. As a result, the (super)gaussian formula for the partition function path integral no longer applies. Instead the EAdS dual partition function is manifestly divergent for $\sigma / \sqrt{N} < -3/4$ and thus predicts a vanishing wave function in this domain \cite{Hertog2011,Conti:2017pqc}. Similar arguments apply to the metric deformations we considered; the dynamical CFT scalar has zero or negative eigenvalues for squashings $\alpha \leq -3/4$ leading to a vanishing wave function in this regime \cite{Conti:2017pqc}.

Finally, for the pseudoscalar deformation, the situation is less clear. The pseudoscalar deformation only affects the eigenvalues of the dynamical CFT spinor, not those of the scalar. Spinors always have negative eigenvalues and these do not pose a problem for the path integral. However, trouble does appear when there are spinor zero eigenmodes resulting in a divergent wave function. The first of these appear when $|m/\sqrt{N}|=3/2$ and thus our results should at most be trusted for $|m/\sqrt{N}|<3/2$.
It would be very interesting to obtain a better understanding of the measure against which the wave function should be evaluated for pseudoscalar bulk fields. To give a spark of hope that the zero-eigenmode divergencies can in due time be dealt with, we point out that an analogous problem appears in the RNS quantization of the superstring. Here, commuting spinor ghosts $\beta$ and $\gamma$ are introduced \cite{Polchinski:1998rr}. These have a number of zero eigenmodes depending on the genus of the worldsheet one consider which produce formal divergencies in the computations, but it is understood how the theory must be modified to resolve these.

For a pseudoscalar bulk deformation, the value of $<N_F>[m]$ can be seen in \cref{pseudoscalarnum}. We stress again that, as discussed in \cref{secwave functionfermion}, this object has been constructed to compute the fermion occupation number at a given value of the bulk scalar, with the latter treated as an external condensate. At large $N$, we see that $N_F$ is zero at $m=0$, that it increases with increasing $|m|$, saturating at complete occupation at $|m/\sqrt{N}| = 3/2$. This agrees beautifully with the earlier bulk computations of \cite{DEath:1986lxx,DEath:1996ejf} as follows: We do not know the bulk action when one integrates out all degrees of freedom, retaining only the bulk pseudoscalar and spin-1/2 field. Still, generically, the leading coupling between a scalar and a spinor is of the Yukawa-type. In the presence of a non-dynamical external scalar field, the Yukawa coupling effectively acts as a mass term for the spinor. We expect the large $N$ limit in the CFT to correspond to the free limit since $N = (G_N \Lambda)^{-1}$ and we work at constant $\Lambda$. One then expects that for a boundary source $m/\sqrt{N}$ and at large $N$, the bulk fermion will effectively behave as a free massive spin-1/2 fermion of mass $m_F = m/\sqrt{N}$. In \cite{DEath:1986lxx,DEath:1996ejf} the Hartle-Hawking wave function,
together with  $<N_F>[m_F]$, was computed for a free fermion in Einstein gravity in de Sitter space. For a massless bulk fermion, they found that $N_F$ should be zero. Next, $N_F$ should increase with increasing $|m_F|$. For modes with $ n + 3/2 \ll |m_F|$ at $l=1$, they found that $N_F$ should saturate at maximum occupation. We are looking at constant spinors with $n=0$, so this completely agrees with our boundary result, within the range of $|m_F|$ where our analysis is to be trusted.

The interplay of the spin-1/2 field with the bulk scalar differs from that with the pseudoscalar, especially in the large $N$ limit. Since $\sum_n a_n = 0$, in the scalar case the part of the wave function with spin-1/2 states excited totally disappears here\footnote{Except at points where there is a competing divergence in $|\Psi |^2$, such as at $\sigma/\sqrt{N} = -9/4$. However we argued these should be at or beyond the edge of the configuration space.}; it appears the scalar and spin-1/2 field do not interact in the large $N$ limit. This may be a consequence of the boundary conditions we employ. The deformations of the boundary CFT correspond to the standard quantisation for the bulk pseudoscalar and to the alternate quantization for the bulk scalar \cite{Hertog:2017ymy}. It would be interesting to explore whether the interplay between the bulk bosons and the spin-1/2 field changes for different boundary conditions.

%%%%%%%%%%%%%%%%
\section{Outlook}
\label{sec:discussion}
%%%%%%%%%%%%%%%%

We have initiated the study of fermionic bulk fields in dS/CFT, working with the dualities relating ${\cal N}=2$ supersymmetric Euclidean vector models with reversed spin-statistics in three dimensions to supersymmetric Vasiliev theories in four-dimensional de Sitter space. Specifically we have holographically evaluated the Hartle - Hawking wave function in the bulk in a number of homogeneous minisuperspace models consisting of a half-integer spin field coupled to a scalar, a pseudoscalar or a metric squashing. 
With a well-motivated measure we have found the wave function to be normalizable and globally peaked at or near the supersymmetric de Sitter vacuum, with a low amplitude for large deformations. A detailed discussion of its behavior and a comparison with earlier bulk computations in the semiclassical limit is given in Section \ref{narrowdiscussion} above.

We have considered the lowest, homogeneous mode of a spin-1/2 bulk field only. It would be natural to look at higher modes, and to study bulk fermions of higher spin. Gravitinos have been studied in a supergravity context in AdS/CFT and in supersymmetric quantum cosmology \cite{Corley:1998qg,Volovich:1998tj,DEath:1988jyp,DEath:1992wve,Moniz:1996pd}. Beyond this, higher spin theories have an infinite tower of fermionic higher spin fields. The expressions for the conserved CFT currents related to these fields are known exactly \cite{Nizami:2013tpa}\footnote{We refer to \cite{Buchbinder:2018nkp,Hutomo:2018tjh,Hutomo:2018iqo} for a recent analysis in $\cN=2$ supersymmetric AdS$_3$ which might be Wick-rotated to supersymmetric $S^3$.}. Since it is relatively straightforward to compute the wave function for fermionic bulk fields (compared to bosonic fields) one might hope to compute the wave function for the lowest modes of these fermionic higher spin fields exactly.

We have seen that the fermionic contributions introduce an interesting $N$-dependence in the theory. dS/CFT at finite $N$ is largely unexplored territory, but recently the Q-model \cite{Anninos:2017eib,Anninos:2019nib} has been put forward as a possible completion of higher-spin dS/CFT. We made use of the measure implied by the Q-model to identify the domain of the wave function in some of the models we considered in this paper. An interesting avenue for future research would be to examine whether fermionic bulk fields can be implemented in the Q-model and to explore how various small $N$ effects play out against each other.

Other possible generalizations in the HS context include the study of fermions in the causal patch version of dS/CFT, currently formulated in terms of a boundary particle mechanics \cite{Neiman:2018ufb}, and in FLRW-like cosmologies in higher spin theory \cite{Aros:2017ror} for which currently a dual description has yet to be found.

Undoubtedly the most challenging open question concerns the formulation of a precise dS/CFT duality in string theory.
One important difference with HS theory, when it comes to dS/CFT, is that string theory has towers of fields of arbitrarily high mass whereas HS theory contains only massless and very light fields. It is well known that light scalar fields in dS with masses below the dS analogue $m^2_{BF}=9/4H^2$ of the Breitenlohner - Freedman (BF) bound behave very differently from more massive fields \cite{Strominger:2001pn,McInnes:2001dq,Hartle:2008ng,Kiritsis:2019wyk,Thomastalk}. It is worth noting in this context that the semiclassical Hartle - Hawking wave function vanishes in dynamical models of de Sitter gravity coupled to massive scalars with masses $m^2 > m^2_{BF}$, at least for reasonably small deformations of de Sitter \cite{Hartle2007}. This suggests that if it computes the Hartle - Hawking wave function, dS/CFT duality should incorporate a final condition on such massive scalars that sets these to zero\footnote{We note, however, that light scalar fields in de Sitter coupled to gravity give rise to the stochastic dynamics associated with eternal inflation and thus come with their own set of challenges \cite{Hawking:2017wrd}.}. This is effectively what we have seen in the minisuperspace models we have analysed, as discussed in Section \ref{narrowdiscussion} above, and this would resolve the tension between dS/CFT and the swampland arguments  \cite{Brennan:2017rbf,Danielsson:2018ztv,Garg:2018reu,Obied:2018sgi,Ooguri:2018wrx,Andriot:2018ept}, some of which suggesting dS with $m^2 > m^2_{BF}$ scalars is unstable\footnote{Whilst the discussion of instabilities has mostly focussed on scalar fields we note that our study of massive spin-1/2 fields implies a similar limit $|m_F| \leq 3/2$, which seems in line with the reasoning above.}.

Finally, it is tempting to speculate that our findings are connected to the supersymmetric dS constructions in exotic string theories \cite{Hull:1998vg}. The latter have vector ghosts in their supergravity limits related to the existence of  non-compact $R$-symmetry groups in their representation of the algebra. However Hull has argued that the massive string states in exotic string theories may well render the de Sitter vacua ghost-free and unitary. In \cite{Hertog:2017ymy} we conjectured that the supersymmetric 
higher-spin theories in dS that we construct are related indeed to the tensionless limit of these exotic string theories. It would be very interesting to explore the extension to dS/CFT of the ABJ triality \cite{Chang:2012kt} linking higher spin theory and string theory in AdS/CFT from a complementary angle and in different regimes.

%%%%%%%%%%%%%%%%
\section*{Acknowledgements}
\noindent
We thank Thomas Van Riet, Nikolay Bobev, Yannick Vreys, Yasha Neiman, Ergin Sezgin, Daniel Grumiller and Dmitry Ponomarev
for helpful discussions. GV thanks HKUST and OIST for their hospitality during part of this work. This work was supported in part 
by the European Research Council grant no. ERC-2013-CoG 616732 HoloQosmos, 
the FWO Odysseus grant G.0.E52.14N, the IAP Programme of the Belgian Science Policy (P7/37)
 and by the KU Leuven C1 grant C16/16/005.
The work of GT-M is also supported by the A. Einstein Center for Fundamental Physics, University of Bern,
and the Australian Research Council Future Fellowship FT180100353.
%%%%%%%%%%%%%%%%

%%%%%%%%%%%%%%%%
\appendix
%%%%%%%%%%%%%%%%

%%%%%%%%%%%%%%%%
\section{Quantization fermions in de Sitter space}
\label{freespinorapp}
%%%%%%%%%%%%%%%%
In this appendix we provide some additional details on the quantization of a free massless spinor in four-dimensional de Sitter space and the definition of $\breve{\psi}$. None of these results are new, we merely summarize some results of \cite{DEath:1986lxx,DEath:1996ejf} without derivation, specialized to the case of a massless two-component spinor and add to this some comments related to higher spin theory. The interested reader can find a more complete treatment in those texts, including also the extension to a massive dirac spinor and the inclusion of dynamical metric and scalar fields. We will consider the metric for a spacetime whose spatial slices are three-spheres,
\begin{equation}
\label{appmetric}
ds^2 = - dt^2 + e^{2 \alpha(t)} d \Omega_3^2 \,,
\end{equation}
with  $d \Omega_3^2$ the metric on the unit three-sphere. We will denote the metric on a spatial slice by $h_{i j}(t)$. When $e^{\alpha(t)} \simeq l \text{ cosh}(t/l)$, one is approximately in a de Sitter regime, with de Sitter radius $l$. One could demand 
this behaviour for the entire spacetime or asymptotically at late times as desired. Our discussion will be valid for generic metrics of the form \cref{appmetric}.

The standard action for a two-component massless spinor $\psi^A$ in tertad $e^a_\mu$ and bispinor for tangent vector $a \rightarrow A A'$ notation has the volume term
\begin{equation}
I_V = - \frac{\ri}{2} \int d^4 x \,e\, \breve{\psi}^{A'} e^\mu_{A A'} D_\mu \psi^A \,.
\end{equation}
Here $\breve{\psi}^{A'}$ is ordinarily the Dirac adjoint $\overline{\psi}^{A'}=(\psi^A)^\dagger$ of $\psi^A$ (but read on!). We follow \cite{DEath:1986lxx,DEath:1996ejf} in introducing a seperate notation to indicate that for, e.g., the purpose of varying the action, computing conjugate momenta, canonically quantizing in an Hamiltinioan approach, one treats $\psi^A$ and $\breve{\psi}^{A'}$ \textit{as if} they were independent. This is the standard textbook approach, though the notation is nonstandard. 

Following this breve notation provides one further advantage in our case. In the supersymmetric de Sitter higher spin theory of \cite{Sezgin:2012ag}, which forms the bulk side of the specific dS/CFT duality treated in this paper, there appears a conjugate, which we will denote with $c$, which acts as the hermitian conjugate $\dagger$ on bosonic quantities. On fermionic quantities, however, it has the special property that $((\psi^A)^c)^c = - \psi^A$. Conjugate spinors in the higher spin theory are defined using this $c$ conjugate rather than the hermitian conjugate. Because we treat $\psi^A$ and $\breve{\psi}^{A'}$ \textit{as if} they were independent, we can treat the cases $\breve{\psi}^{A'}=(\psi^A)^\dagger$ and $\breve{\psi}^{A'}=(\psi^A)^c$ at the same time and the need to distinguish the two will not come up in this appendix or \cref{secwave functionfermion}. We can thus discuss the general framework for including fermions in dS/CFT in both ordinary gravity + matter theories and higher spin theories at the same time.

In the presence of an initial $\Sigma_I$ and/or final $\Sigma_F$ boundary surface, the action also requires boundary terms\footnote{When computing the Hartle-Hawking state, there would be only a final surface but no initial one. Further, the metric would also be dynamical and allowed to take complex values. Here, we discuss the free fermion generically.}
\begin{equation}
I_B = \frac{\ri}{2}\left(\int_{\Sigma_I}+\int_{\Sigma_F}\right) d^3 x \,h^{1/2}\,\breve{\psi}^{A'} n_{A A'} \psi^A \,,
\end{equation}
with $n^{A A'}$ the unit timelike vector normal to the surface. The conjugate momentum to $\psi^A$ is given by
\begin{equation}
\pi_A = - \frac{\ri}{2} h^{1/2} n_{A A'} \breve{\psi}^{A'} \,.
\end{equation}
Using this identity, we can and will take the fundamental dynamical variables of the theory to be $\psi^A$ and $\breve{\psi}^A$ rather than $\psi^A$ and $\pi_A$.

One finds the classical Dirac bracket
\begin{equation}
\label{diracbracket}
\{ h^{1/4} \psi^A(x) \,, h^{1/4} \breve{\psi}^{A'}(x') \}_D = -2 \ri n^{A A'} \delta(x,x')  \,,
\end{equation}
and all other Dirac brackets zero. One quantizes by the usual procedure, turning the Dirac brackets into anticommutators with factor of $\ri \hbar$ inserted. One can expand spinor-valued functions in the background of \cref{appmetric} in spatial three-sphere harmonics. We denote the positive frequency three-sphere harmonics by $\rho^{np}_A$ and $\sigma^{n p}_{A'}$; the negative frequency harmonics by $\overline{\rho}^{np}_{A'}$ and $\overline{\sigma}^{n p}_{A}$. The indices run over $n=0,1,2,...$ and $p=1,2,..,(n+1)(n+2)$.
The expansions of $\psi^A$ and $\breve{\psi}^{A'}$ are then given by\footnote{This expansion is completely general and valid. However, in practical bulk computations it is more convenient to expand e.g. as $\alpha_n^{p q}s_{n q }\rho^n_{p A}$ instead of $s_{n p} \rho^{n p}$, with the $\alpha_n^{n p}$ constant matrices auspiciously chosen to ensure there are no couplings between different terms with the same value of $n$ when filling these expansions back into the action. These issues related to the degeneracy are not relevant in our context as we focus only on one of the two $n=0$ modes.}
\bsubeq
\begin{align}
\psi_A &= \frac{e^{- 3 \alpha / 2}}{2 \pi} \sum_n \sum_p \big( s_{n p}(t) \rho^{n p}_A + \breve{t}_{n p}(t) \overline{\sigma}^{n p}_A \big) \,, \\
\breve{\psi}_{A'} &= \frac{e^{- 3 \alpha / 2}}{2 \pi} \sum_n \sum_p \big( \breve{s}_{n p}(t) \overline{\rho}^{n p}_{A'} + t_{n p}(t) \sigma^{n p}_{A'} \big) \,.
\end{align}
\esubeq
The dynamical coeffecients $s_{n p}(t)$, $\breve{t}_{n p}(t)$, $\breve{s}_{n p}(t)$, $t_{n p}(t)$ are Grassmann-valued for anticommuting spinors, as is the case here.
The prefactors in front of these expansions could equally well have been absorbed into the dynamical  coefficients. This choice of prefactors is convenient as they counteract the factors of $h^{1/4}$ appearing in \cref{diracbracket}. This ensures that upon quantization one obtains \cref{anncreat} without any further factors or metric dependence.
%%%%%%%%%%%%%%%%
\bibliographystyle{utphys}
\bibliography{dSCFT}
\end{document}